\documentclass[a4paper,fleqn,usenatbib]{mnras}

\usepackage{txfonts}

\usepackage[T1]{fontenc}
\usepackage{ae,aecompl}

\usepackage{graphicx}	
\usepackage{pdflscape}
\usepackage[maxfloats=25]{morefloats}

\title[Polarisation of OH in planetary nebulae]{Polarisation properties of OH emission in planetary nebulae}

\author[G\'omez et al.]{
Jos\'e F. G\'omez,$^{1}$\thanks{E-mail: jfg@iaa.es}
Lucero Uscanga,$^{2}$
James A. Green,$^{3}$
Luis F. Miranda,$^{1}$
\newauthor
Olga Su\'arez$^{4}$
and
Philippe Bendjoya$^{4}$
\\
$^{1}$Instituto de Astrof\'{\i}sica de Andaluc\'{\i}a, CSIC, Glorieta de la Astronom\'{\i}a s/n, 18008 Granada, Spain\\
$^{2}$Departamento de Astronom\'{\i}a, Universidad de Guanajuato, A.P. 144, 36000 Guanajuato, Gto., Mexico\\
$^{3}$Australia Telescope National Facility, CSIRO Astronomy and Space Science, P.O. Box 76, Epping, NSW 1710, Australia\\
$^{4}$Laboratoire Lagrange, UMR 7293, Universit\'e de Nice Sophia-Antipolis, CNRS, Observatoire de la C\^ote d'Azur, F-06304 Nice, France}

\date{Accepted XXX. Received YYY; in original form ZZZ}

\pubyear{2016}

\begin{document}
\label{firstpage}
\pagerange{\pageref{firstpage}--\pageref{lastpage}}
\maketitle

\begin{abstract}
We present the interferometric, full-polarisation observations of the four ground-state transitions of OH, toward five confirmed and one candidate OH-emitting planetary nebulae (OHPNe). OHPNe are believed to be very young PNe, and information on their magnetic fields (provided by their polarisation) could be key to understand the early evolution of PNe. We detect significant circular and linear polarisation in four and two objects, respectively. Possible Zeeman pairs are seen in JaSt 23 and IRAS 17393-2727, resulting in estimates of magnetic field strengths between 0.8 and 24 mG.
We also report the new detection of OH emission at 1720 MHz toward Vy 2-2, making it the third known PN with this type of emission. { We suggest that younger PNe have spectra dominated by narrow maser features and higher degrees of polarisation}. Shock-excited emission at 1720 MHz seems to be more common in PNe than in early evolutionary phases, and could be related to equatorial ejections during the early PN phase.

\end{abstract}




\section{Introduction}

\label{sec:intro}

Planetary nebulae (PNe) are one of the last phases in the evolution of low and intermediate mass stars
($0.8- 8$ M$_\odot$). Their immediate precursors are stars in the asymptotic giant branch (AGB), characterized by
a strong mass-loss  \citep[with rates up to $10^{-4}$  M$_\odot$ yr$^{-1}$,][]{blo95a}, followed by a short  { \citep[$\simeq 10^2 -10^4$ yr,][]{blo95b}} transitional 
post-AGB phase. While the morphology of the mass-loss processes in the AGB is mainly spherical, PNe
usually show complex bipolar and multipolar structures. It is thought that the shaping of these asymmetrical PNe is determined by highly
collimated jets ejected during the post-AGB and early PN phases, which open cavities in the previously
ejected AGB envelope \citep{sah98}. However, the ultimate powering agent of these jets and their timescale of action are
still largely unknown. A reasonable assumption is that jets in post-AGBs and young PNe are launched and collimated by magnetic fields { \citep*[e.g.,][]{gar05,nor06,toc14}}, 
likely requiring the presence of binary central stars { \citep{sok98,dem09,gar14,sok14}}. 

OH masers are potentially important tools in the study of the transition to PNe, due to their timescales
of pumping in evolved objects. OH masers are expected to disappear $\simeq 1000$ years after the end of the
AGB mass-loss \citep*{lew89,gom90}. Considering that the post-AGB phase lasts $\simeq 10^2 - 10^4$
yr  { (depending on the stellar mass)}, 
PNe showing OH emission are expected to be very young.
Very few members of this special type of OH-emitting PNe (OHPNe) are known: \citep{zij89}
initially cataloged 12 possible OHPNe, mainly based on single-dish OH data, while interferometric data
\citep{usc12} confirmed only 6 sources as bona fide members of this class. A seventh OHPN has been recently confirmed  \citep[IRAS 16333-4807,][]{qia16}.
These objects could be among the youngest PNe, and therefore, key objects in the study of PN shaping.
Moreover, the presence of OH masers is important from a practical point of view, since their polarimetric
observation can provide information on the magnetic field strength, by measuring Zeeman splitting. 
The presence of magnetic fields would cause the two sigma-components of the OH lines, seen to be circularly polarised when the field is not purely in the plane of the sky, to be shifted alternately in frequency 
\citep[see][for a recent review in the context of maser emission]{gre14}, { in what is known as a ``Zeeman pair''}. 
{ This splitting typically results in a characteristic S-shaped profile in spectra of the Stokes V parameter (with a negative and a positive peak close together in frequency)}. Through examining this profile and its characteristics for multiple sources, it is then possible to study whether magnetic fields play a significant role in the shaping of PNe. Magnetic field strengths have been estimated for several AGB and post-AGB stars using polarimetric maser observations of { OH, H$_2$O, and SiO \citep{vle12}, showing ordered magnetic fields of a few G close to the central star, and decreasing up to $\simeq 1$ mG at 1000 au. However,} 
such observations are scarce in PNe, given the low detection rates of maser emission in this type of objects.

OHPNe do not seem to form an homogeneous group \citep{usc12}. While objects such as
IRAS 17347$-$3139 or K3-35 are optically obscured and/or compact in size  \citep[$\simeq 10000$ AU,][]{deg04,bla14}, others like NGC 6302 are optically bright and very extended  \citep[$\simeq 1$ pc,][]{szy11}, and thus, they are probably more evolved PNe. 
Therefore not all OHPNe seem to strictly conform to their expected nature as extremely young PNe, and
some other process may sustain OH emission for a longer time than previously assumed. This apparent
age spread within the OHPN group is interesting, since it would allow us to study the evolution of their
properties, and to derive information about the shaping of PNe. In particular, if magnetic fields are indeed
a key factor in the powering and collimation of jets leading to asymmetrical PNe, we would expect these
fields to be significantly stronger in obscured and/or compact (thus probably younger) sources than in the more evolved ones. 

Previous polarimetric studies of OH-emitting evolved objects with single dish telescopes \citep*{szy04,wol12,gon14} included OHPNe among their targets. These studies showed that,
while circularly polarised OH features are common in OHPNe, the detection of Zeeman pairs is far more elusive. Interferometric observations have shown more promising candidates. The only OHPN in which Zeeman splitting has been  distinctively detected is IRAS 16333-4807 at 1720 MHz \citep{qia16}, with a derived magnetic field strength of $\simeq 11$ mG. Another possible Zeeman pair has been reported in K 3-35, at 1665 MHz \citep{gom09}, giving an estimate of $B\simeq 0.9$ mG. Both objects are probably among the youngest PNe, as they are both OH and H$_2$O maser emitters \citep{gom15}. 

In this paper we present interferometric, full-polarisation
observations of most of the known OHPNe, in order to study the characteristics of their magnetic fields.

\section{Observations and data analysis}

\label{sec:obs}
Observations were carried out with the Australia Telescope Compact
Array (ATCA), in its 6A configuration \citep{ste15}. We used the
Compact Array Broadband Backend (CABB) in its 1M-0.5k mode, with 2
intermediate frequency bands (IFs), each with a 2 GHz bandwidth
sampled over 2048 broadband channels of 1 MHz. We obtained data in
full polarisation (two linear polarisations and their corresponding
cross-polarisations, so as to form the full Stokes parameters, I, Q, U
and V). Both IFs were centered at 2100 MHz, providing redundant
data. We used the first IF in our data processing, except when noted
otherwise. Twelve 1MHz broadband channels (four groups of 3 channels)
were zoomed in, to provide a finer spectral resolution of 0.5 kHz to
observe spectral lines. With this setup, we observed the OH
ground-level transitions of rest frequencies 1612.2309, 1665.4018,
1667.3590, and 1720.5299 MHz, with a total bandwidth of 2 MHz each,
sampled over 4096 channels of 0.5 kHz (total velocity coverage and
resolution of 381 and 0.09 km s$^{-1}$ at 1612 MHz, respectively). We
observed a total of 5 fields. The source PKS B1934-638 was used to set the absolute flux scale and to calibrate the spectral response of the instrument. Table \ref{tab:obs} gives a summary of some additional observational parameters. The velocities in our data are given to respect to the kinematic definition of the Local Standard of Rest (LSRK).

\begin{table*}
	\centering
	\caption{Observational parameters.}
	\label{tab:obs}
	\begin{tabular}{llllllr} %
		\hline
		Target & \multicolumn{2}{c}{Phase centre} &  Gain calibrator & Date & $\theta_{1612}$$^a$ & p.a.$^ b$\\
			  &	R.A.(J2000)			& 	Dec(J2000)		& 	&  & (arcsec) & (deg)\\
		\hline
		NGC 6302	  & 17:13:44.47 & $-37$:06:11.6 & PMN J1733$-$3722 & 2012-AUG-14 & $11.8\times 6.1$ & $-18$ \\
		JaSt 23	  & 17:40:23.08 & $-27$:49:12.3 & PMN J1713$-$2658 & 2012-AUG-15 & $15.9\times 5.6$ & 0\\
		IRAS 17393-2727 & 17:42:33.14 & $-27$:28:24.7 & PMN J1713$-$2658 & 2012-AUG-16 & $21.0\times 6.8$ & $-38$\\
		Vy 2-2 & 19:24:22.22 & +09:53:56.3 & PMN J1938+0448 & 2012-AUG-16 & $39.5\times 6.2$ & $-1$\\
		K 3-35 & 19:27:44.03 & +21:30:03.6 &  PKS J1925+2106 & 2012-AUG-16 & $24.2\times 5.9$ & 0\\
		\hline
	\end{tabular}\\
	$^ a$ Full width at half maximum of the synthesized beam for the images of maser emission at 1612 MHz. The beam size for the other maser transitions scales as $\theta\propto \nu^{-1}$, where $\nu$ is the rest frequency. The beam parameters for the continuum emission are given in the corresponding figures.\\
	$^ b$ Position angle (north to east) of the synthesized beam of maser emission.
\end{table*}

Our main targets were five OHPNe, which represent all confirmed sources of this class \citep{usc12}, except IRAS 17347-3139, whose intensity and polarisation levels \citep{taf09} indicated that we would not have enough sensitivity to detect polarised emission, and IRAS 16333-4807, which was confirmed as a OHPN after our observations were carried out \citep{qia16}. Besides these five main targets, the OHPN candidate IRAS 17375-2759 \citep[with known OH emission, but whose nature as a PN is not confirmed,][]{usc12} was within the primary beam of the observations toward JaSt 23, and its OH emission has also been processed, and presented in the following sections. These observations are the first full-polarimetric OH observations of these objects carried out with an interferometer.

We calibrated and processed the data using the {\sc miriad} software. When creating images, visibilities were weighted with a robust parameter of 0.5 (as defined in task {\sc invert}) and deconvolved with the {\sc clean} algorithm. For the broadband (continuum data) we first flagged out the channels in which OH line emission could appear, and obtained maps performing frequency synthesis using the whole 2 GHz band (covering a frequency range of 1100-3100 MHz). OH spectra of Stokes I, Q, U, and V were obtained with task uvspec, applying a spatial shift corresponding to the distance from the phase centre to the peak of the OH emission determined from images. 
With the ATCA having native linear feeds, and adopting the standard optical definition for the intensity of circularly polarised emission ($I={\rm RCP}+{\rm LCP}$, where RCP and LCP stand for right and left circular polarisations, respectively), our spectra of individual circular polarisations were obtained as ${\rm RCP} = (I+V)/2$ and ${\rm LCP} = (I-V)/2$. Note that this definition differs from that sometimes used in the literature (${\rm RCP} = I+V$ and ${\rm LCP} = I-V$), especially with data taken from radio telescopes with native circular feeds. In those cases, the previously published LCP and RCP spectra should be divided by two, for a proper comparison with ours.  The spectrum of the linear polarisation { was obtained as} ${\rm LINP} = \sqrt{Q^2+U^2}$. Fractional polarisation was estimated as $p_c=V/I$, $p_l={\rm LINP}/I$. { The fractional linear polarisation ($p_l$) and its uncertainty were corrected by the noise bias using the Modified Asymptotic estimator of \cite{pla14}, assuming no correlation between the noise in Q and U. However, given the high signal-to-noise ratio of the OH features with detected linear polarisation, this bias correction was small. We also determined the position angle of the linear polarisation as $\psi = 0.5\arctan(U/Q)$.}

Uncertainties quoted throughout this paper are at a $2\sigma$ level. The uncertainties in flux density only take into account the noise in the images, and not the possible errors in the determination of the absolute flux density scale. Likewise, the positional uncertainties given here must be interpreted as relative errors between sources in each map (or between spectral components in the same source), and not absolute astrometric errors (which are typically on the order of 10\% of the synthesized beam size). { Upper limits are given at a 3$\sigma$ level. In the case of the lineal polarisation, in which the noise induces a positive bias, upper limits are calculated as a 3$\sigma$ level above the mean of the linear polarisation spectrum.}

\section{Results}

\label{sec:results}

\subsection{Overview}

\subsubsection{Radio continuum}

\begin{figure}
	\includegraphics[width=0.9\columnwidth]{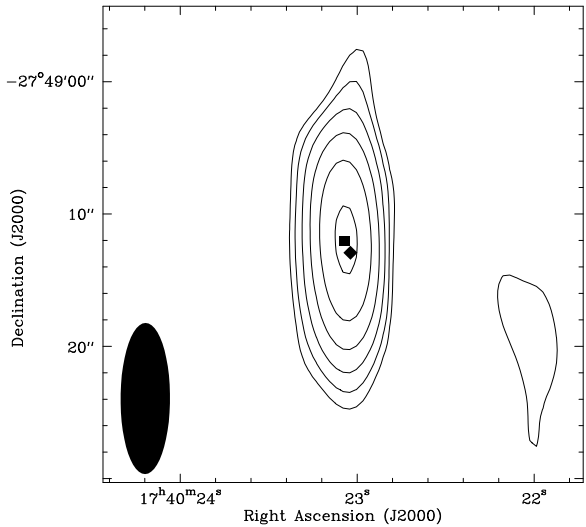}\\
		\caption{Contour map of the continuum emission in JaSt 23. The contour levels are $3\times 2^n$ times 15 $\mu$Jy beam$^{-1}$ (the rms of the map), with $n=0$ to 5. The black square and diamond mark the  location of peak emission at 1612 and 1665 MHz, respectively. The filled ellipse represents the synthesized beam of the continuum image ($11.4''\times 3.7''$, p.a. $=0^\circ$)}

	\label{fig:jast_cont}
\end{figure}

\begin{figure}
	\includegraphics[width=0.9\columnwidth]{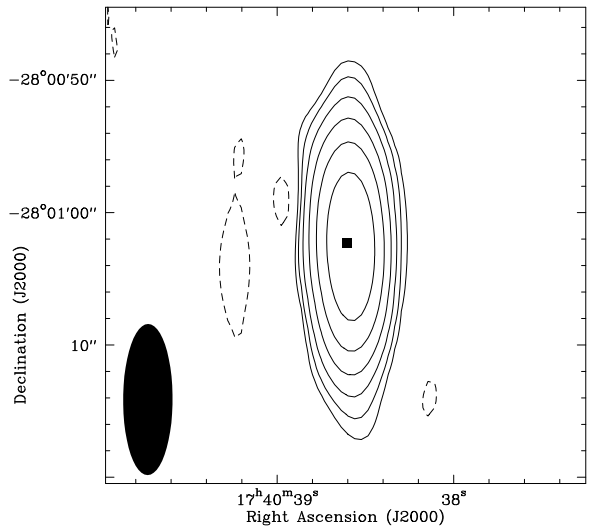}\\
		\caption{Contour map of the continuum emission in IRAS 17375-2759. The contour levels are $3\times 2^n$ times 27 $\mu$Jy beam$^{-1}$ (the rms of the map), with $n=0$ to 5. The black square  marks the  location of peak emission at 1612 MHz. The other maser components appear at the same location in the image. The filled ellipse represents the synthesized beam of the continuum image ($11.4''\times 3.7''$, p.a. $=0^\circ$).}
	\label{fig:iras0_cont}
\end{figure}

\begin{figure}
	\includegraphics[width=0.9\columnwidth]{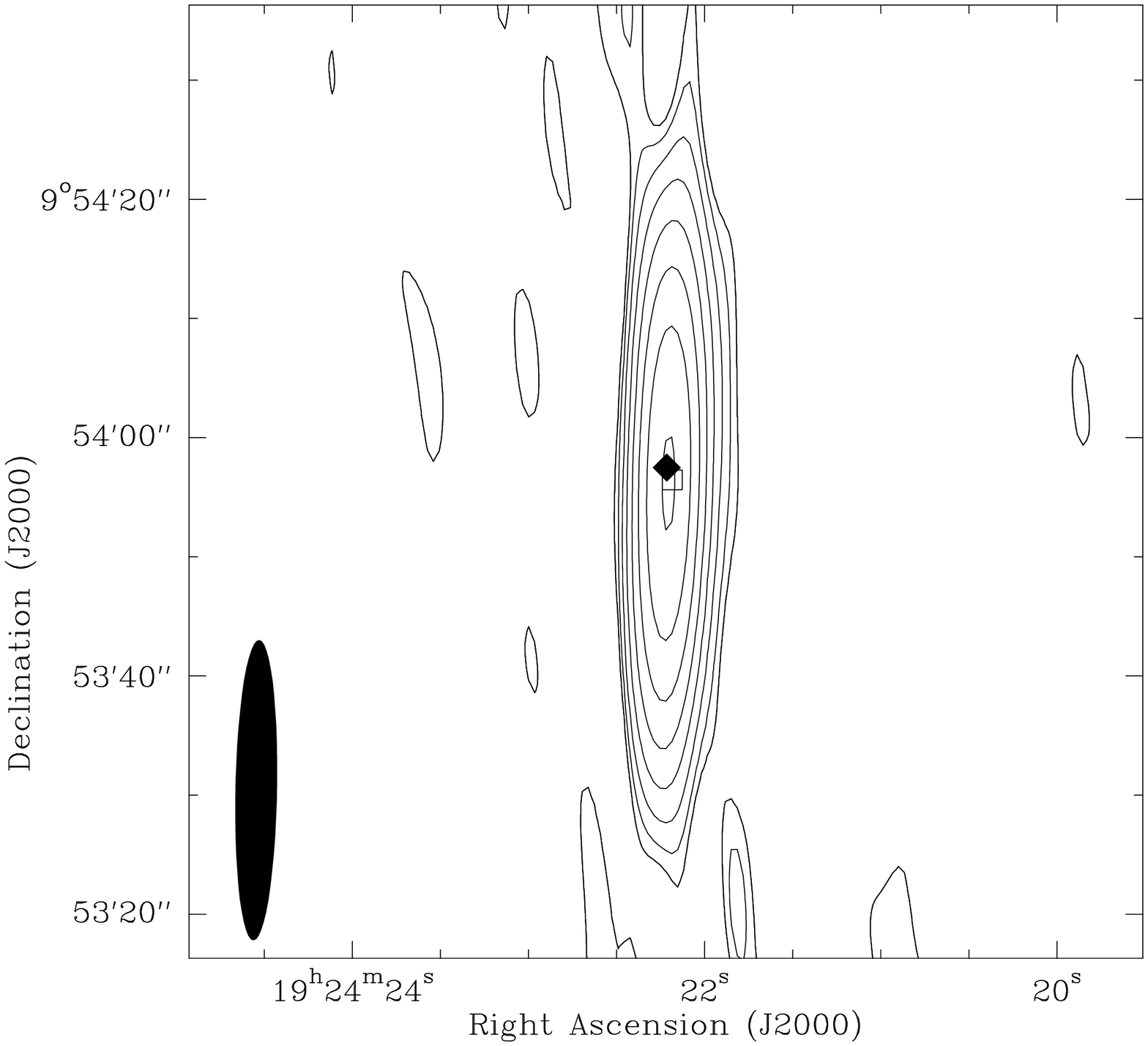}\\
	\caption{Contour map of the continuum emission in Vy 2-2. The countour levels are $3\times 2^n$ times 40 
	$\mu$Jy beam$^{-1}$ (the rms of the map). The open square and filled diamond mark the location of the emission at 1612 and 1720 MHz, respectively. The filled ellipse represents the synthesized beam of the continuum image ($25.2''\times 3.5''$, p.a. $=1^\circ$)}
	\label{fig:vy_cont}
\end{figure}

Radio continuum emission is detected in all six sources. The emission is unresolved, except in the case of NGC 6302, which shows clear extended lobes. 
Table \ref{tab:cont} shows the parameters of the continuum emission, and images are presented in Figs.\ \ref{fig:ngc_cont}, \ref{fig:jast_cont}, \ref{fig:iras0_cont}, \ref{fig:iras_cont}, \ref{fig:vy_cont}, and \ref{fig:k3_cont}.

\begin{table*}
	\centering
	\caption{Parameters of radio continuum emission at 2100 MHz}
	\label{tab:cont}
	\begin{tabular}{lllll} %
		\hline
		Source & R.A.(J2000) & Dec(J2000) &  Error$^a$ & $S_\nu$$^b$ \\
		  &				& 			& (arcsec)	& (mJy) \\
		\hline
		NGC 6302	  & 17:13:44.489 & $-37$:06:11.71 & (0.17, 0.23) & $2459\pm 4$\\
		JaSt 23 	  & 17:40:23.067 & $-27$:49:12.00 & (0.08, 0.26) & $1.51\pm 0.06$\\
		IRAS 17375-2759 & 17:40:38.584 & $-28$:01:02.53 & (0.04, 0.12) & $4.78\pm 0.15$\\
		IRAS 17393-2727 & 17:42:33.116 & $-27$:28:25.2 & (0.23, 0.3) & $0.86\pm 0.10$\\
		Vy 2-2			& 19:24:22.204 & $+09$:53:56.2 & (0.04, 0.3) & $16.53 \pm 0.24$\\
		K 3-35			& 19:27:44.029 & $+21$:30:03.5 & (0.03, 0.15) &  $11.9\pm 0.3$\\
		\hline
	\end{tabular}\\
	$^a$Positional error in right ascension and declination\\
	$^b$Flux density
\end{table*}

\subsubsection{OH emission}

\begin{table*}
	\centering
	\caption{Main parameters of OH lines}
	\label{tab:oh}
	\begin{tabular}{lllllrr} %
		\hline
		Source & Transition &  $V_{\rm peak}$$^a$ &  $V_{\rm min}$$^b$ & $V_{\rm max}$$^c$ & $S_{\rm peak}$$^d$ & $\int S_\nu dv$$^e$ \\
		  &		(MHz)		& 	(km s$^{-1}$) &	(km s$^{-1}$)	& (km s$^{-1}$)	& (Jy) & (Jy km s$^{-1}$)\\
		\hline
		NGC 6302 & 1612 & $-41.5$ & $-54.7$ & $-32.0$ & $0.586\pm 0.026$ & $4.16\pm 0.04$\\
				& 1665 & $-43.0$ & $-59.5$ & $-36.4$ & $-0.097\pm 0.026$ & $-0.82\pm 0.03$\\ 
				& 1667 & $-45.3$ & $-55.4$ & $-33.6$ & $0.148\pm 0.026$& $0.42\pm 0.04$ \\
				& 1720 & $-39.9$ & $-54.2$ & $-34.2$ & $-0.159\pm 0.026$ & $-1.71\pm 0.03$\\
		JaSt 23 & 1612 & $115.7$ & $+104.6$ & $+122.6$ & $1.36\pm 0.04$ & $4.91\pm 0.02$\\
				& 1665 & $118.2$ & $+103.3$ & $+123.9$ & $0.30\pm 0.03$ & $2.86\pm 0.03$\\
				& 1667 &         &         &         & $<0.04$\\
				& 1720 &         &         &         & $<0.04$\\
		IRAS 17375-2759 & 	1612 & 	$+24.1$ & $+18.6$ & $+32.9$  & $0.87\pm 0.04$ & $4.59\pm 0.05$\\
						& 1665	 & 			&		  &			& $<0.07$ \\
						& 1667   & $+19.1$ & $+7.2$ & $+32.8$	& $0.38\pm 0.04$ & $3.11\pm 0.07$\\
						& 1720	 &			&		&			& $<0.07$ \\
		IRAS 17393-2727 & 1612 & $-123.2$ & $-130.3$  & $-89.6$ & $73.7  \pm 1.2$ & $235.24\pm 0.06$\\
						& 1665 & $-116.4$ &  $-124.8$ & $-101.6$ & $0.85\pm 0.05$ & $3.48\pm 0.06$\\
						& 1667 & $-124.5$ &  $-145.6$ & $-110.3$ & $1.23 \pm 0.05$ & $5.76\pm 0.09$\\
						& 1720 &		&			&			& $<0.07$\\
		Vy 2-2			& 1612	& $-61.2$ & $-64.6$	&	$-55.4$	& $4.29\pm 0.06$ & $16.78\pm 0.05$ \\
						& 1665	&			&		&			& $<0.10$\\
						& 1667	&			&		&			& $<0.10$\\
						& 1720  & 	$-54.8$	& $-56.2$  & $-54.0$ &  $0.15\pm 0.04$ & $0.21\pm 0.06$\\
		K 3-35			& 1612	&  $+9.2$	& $-5.1$ & $+22.4$  & $0.87\pm 0.03$ & $2.22\pm 0.10$\\
						& 1665	 &  $+17.4$ & $+16.4$ &	$+23.0$	& $0.08\pm 0.03$ & $0.21\pm 0.06$\\
						& 1667  &			&			&		& $<0.07$ \\
						& 1720  &$+21.2$	& $+20.6$	& $+22.2$ & $0.21\pm 0.04$ & $0.21\pm 0.03$\\ 
		\hline
	\end{tabular}\\
	$^a$LSRK velocity of the peak line emission.\\
	$^b$Minimum LSRK velocity at which emission is detected.\\
	$^c$Maximum LSRK velocity at which emission is detected.\\
	$^d$Peak flux density.\\
	$^e$Velocity-integrated flux density.
\end{table*}

OH emission at 1612 MHz is also detected in all sources (Table \ref{tab:oh}). This transition is the strongest of the observed OH lines in all cases, as is usually the case in evolved stars. Polarised emission is detected in most  of them (the only exceptions are NGC 6302 and  IRAS 17375-2759). In some cases, the spectra of the Stokes V parameter shows an S-shape { that may indicate the presence of a Zeeman pair}. However, the analysis of these spectra is complicated, since most of the sources shows wide spectra, probably the result of several blended OH components. These different components could arise from different zones in the circumstellar envelope and thus, have different polarisation behaviours.  Obviously, Zeeman pairs are more easily determined when individual velocity features  can be identified in the spectrum \citep[as in the case of IRAS 16333-4807,][]{qia16}, but in our sources only K 3-35 shows a spectrum with relatively narrow and distinct components.

When the Zeeman splitting is larger than the { Doppler} linewidth of the individual components, the magnetic field can be readily obtained by measuring the velocity shift ($\Delta v$) between the positive and negative peaks { in the spectrum} of the Stokes-V parameter, with

\begin{equation}
\Delta v = A_v B \cos (\theta),
\end{equation}
where $A_v$ is the coefficient that quantifies the Zeeman splitting 
\citep[with values 0.123, 0.590, 0.354, 0.113 km s$^{-1}$ mG$^{-1}$ for the transitions at 1612, 1665, 1667, and 1720 MHz, respectively;][]{pal67,coo75}, B is the magnetic field, and $\theta$ is the angle between the magnetic field and the line of sight. 

However, if the velocity shift is smaller than the Doppler linewidth, the magnetic field can be estimated by fitting the V spectra as \citep[see, e.g.,][]{eli98,mod05}: 
\begin{equation}
V(v) = c_1 I(v) + c_2 \frac{dI}{dv} 
\label{eq:fit}
\end{equation}
in the  case of thermal emission or saturated masers or
\begin{equation}
V(v) = c_1 I(v) + \frac{c_2 \frac{dI}{dv}}{c_3 + \ln\left[\frac{I(v)}{I(v_0)}\right]}
\label{eq:fit_unsaturated}
\end{equation}
for unsaturated masers, 
where $V(v)$ and $I(v)$ are the spectra of the V and I Stokes parameters, respectively, $I(v_0)$ is the intensity of Stokes-I at the line centre, and $c_1$, $c_2$, $c_3$ are the parameters which are varied in the fit. The { parameter $c_2$ gives} the measure of Zeeman splitting, and is directly related to the magnetic field,
\begin{equation}
c_2 = \frac{k}{2} A_v B \cos (\theta),
\label{eq:b}
\end{equation}
where k is a constant that depends on the emission mechanism involved \citep{eli98}, and is $k=1$ for thermal emission, and $8/3p$ for maser emission, with $p=1$ in the unsaturated regime, while for saturated masers it is 1 for filaments, 2 for planar masers, and 3 for three-dimensional masers.
Thus, with these fits, we  can estimate the projection of the magnetic field along the line of sight ($B\cos(\theta)$). We tried this fitting to all spectra of Stokes-V in our sources. None gave meaningful results under the assumption of unsaturated maser emission (eq. \ref{eq:fit_unsaturated}), while reasonable fits where obtained for JaSt 23 in the saturated regime (eq. \ref{eq:fit}).
The absence of pure Zeeman pairs in the rest of the sources can be explained { if there are significant variations in the magnetic field strength throughout the OH-emitting regions}. In this case, not all the OH emission undergoes the same Zeeman splitting, and the fitting of Equation \ref{eq:fit} cannot provide good results, as the derivative of the intensity is not a good estimate of the Stokes-V spectrum.

Even for JaSt 23, the results should be taken with care, as there is no guarantee that the S-shaped V-spectra are actually { indicating} Zeeman pairs, but could be the results of separate areas of gas with different amounts of circular polarisation. These possible Zeeman pairs should be observed with even higher angular resolution, to separate the different areas where the OH emission arises from, and to ascertain that the negative and positive peaks of the Stokes-V spectrum are at the same position.

We present the general parameters of the OH emission in these sources in Table \ref{tab:oh}. In the following we discuss each source in detail.

\subsection{NGC 6302}

\subsubsection{Radio continuum}

\begin{figure}
	\includegraphics[width=0.88\columnwidth]{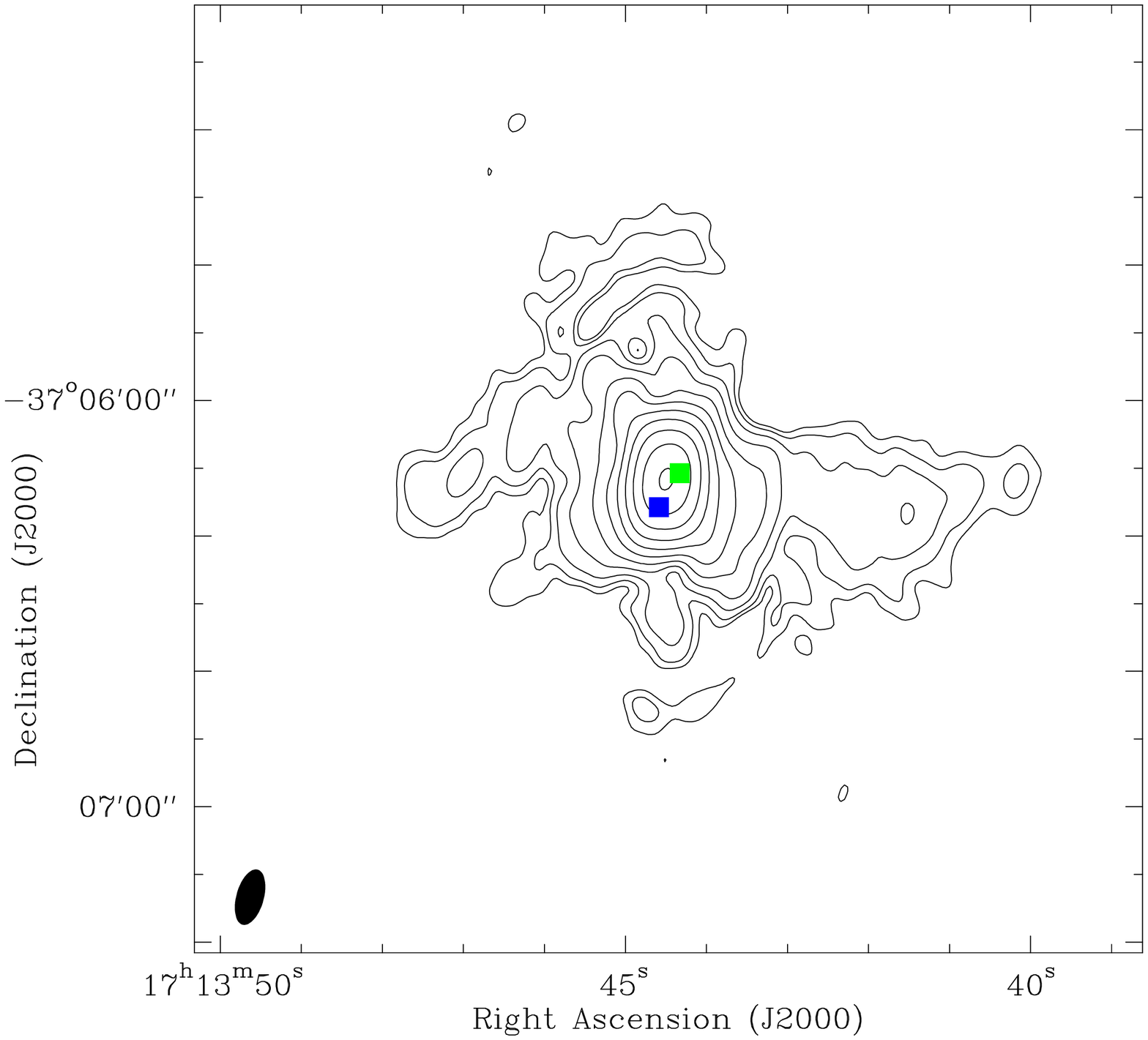}\\
	\includegraphics[width=0.95\columnwidth]{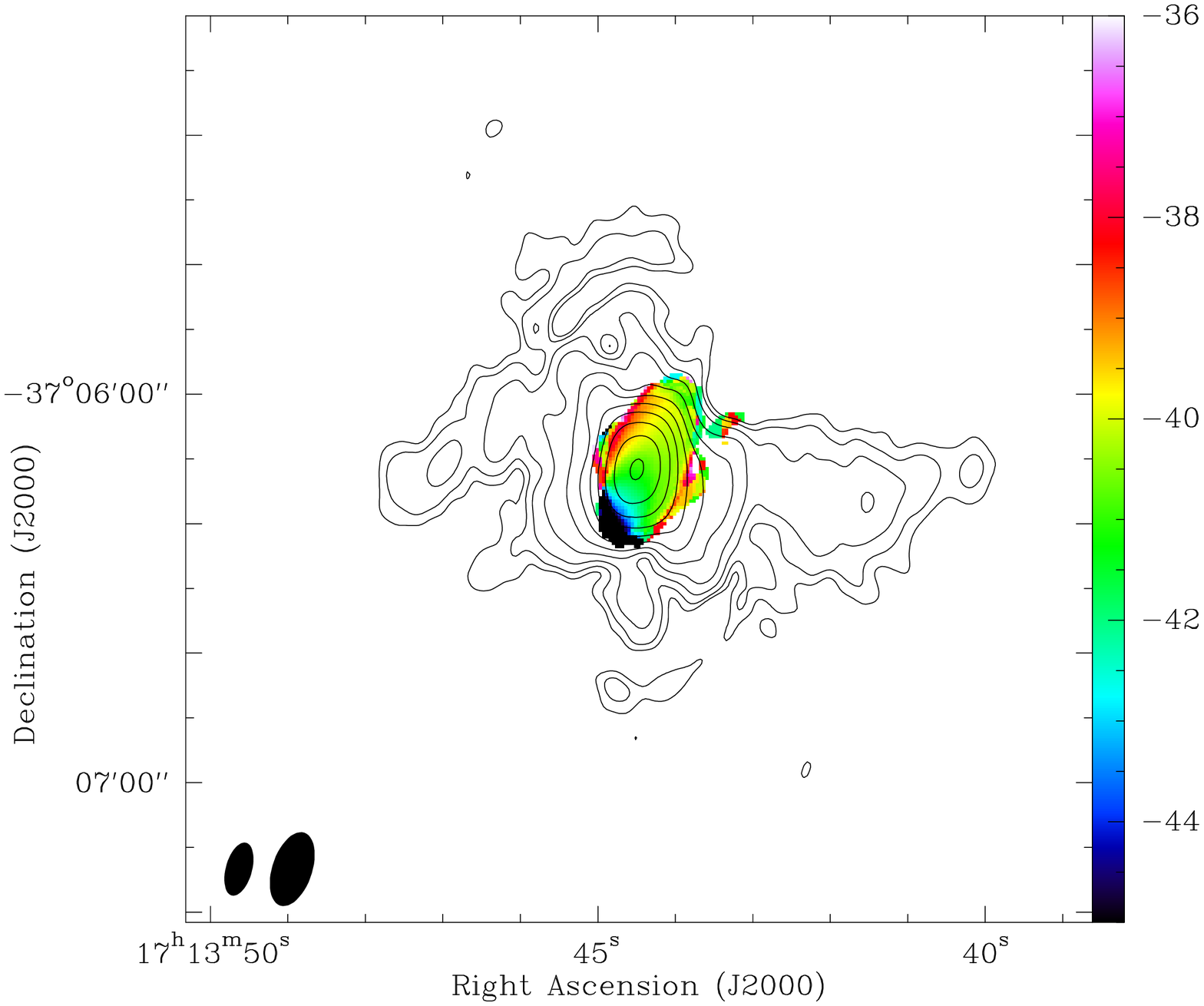}
	\caption{(Top) Contour map of the continuum emission in NGC 6302. The contour levels are $3\times 2^n$ times 0.3 mJy beam$^{-1}$ (the rms of the map), with $n=0$ to 10. Green and blue filled squares represent the location of the distinct OH components at 1612 MHz, ($-41.5$ and $-51.4$ km s$^{-1}$, respectively). (Bottom): Contour map of the continuum emission and and colour map of moment 1 (mean velocity) of the OH emission at 1612 MHz. The colour scale at the right represents the mean velocity in km s$^{-1}$. The filled ellipses represent the synthesized beams of the continuum ($8.4''\times 4.0''$, p.a. $=-15^\circ$) and the OH transition at 1612 MHz ($11.8''\times 6.1''$, p.a. $-18^\circ$). }
	\label{fig:ngc_cont}
\end{figure}

The morphology of the continuum emission (Fig. \ref{fig:ngc_cont}) consists of a bright central part, with a deconvolved size of $6.9''\times 5.4''$, p.a. $=24^\circ$, and  a pair of weaker lobes along the E-W direction, the same direction as the large optical nebula. The western lobe is more extended, with a size of $\simeq 1'$. 
The arc-like features seen at low level $\simeq 35''$ north and south of the continuum peak are likely to be artifacts from the point-spread function of the observation that could not be properly deconvolved.
The lobes seen in our map resemble those seen in the low-resolution map (beam $\simeq 12''\times 8''$) of \cite{gom89} at 4.9 GHz. We do not resolve the ionized torus in the higher resolution map of \cite{gom89}, which has a size of $\simeq 5.0''\times 2.2''$ and a position angle $\simeq -30^\circ$. The bright central emission in our map (p.a. $=24^\circ$) could be related to the NE-SW extension seen by \cite*{gom93} at 8.4 GHz. The flux density we measured is consistent with the reported values of 1.9 Jy \citep[at 1.4 GHz,][]{con98} and 2.4-3.4 Jy \citep[at 4.9 GHz][]{gom89,wri94}.

\subsubsection{OH lines}

\begin{figure*}
	\includegraphics[width=1.1\columnwidth]{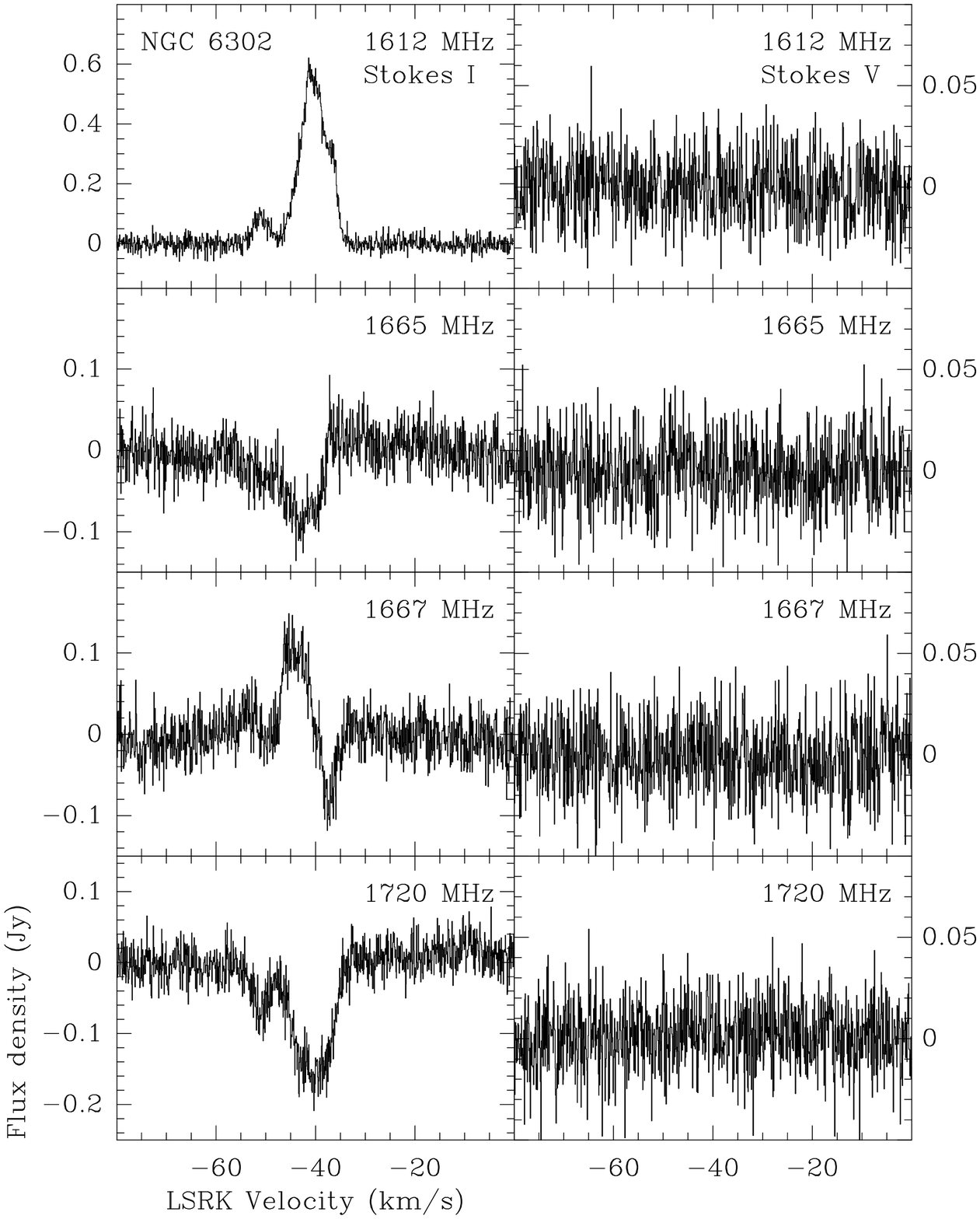}
		\includegraphics[width=0.8\columnwidth]{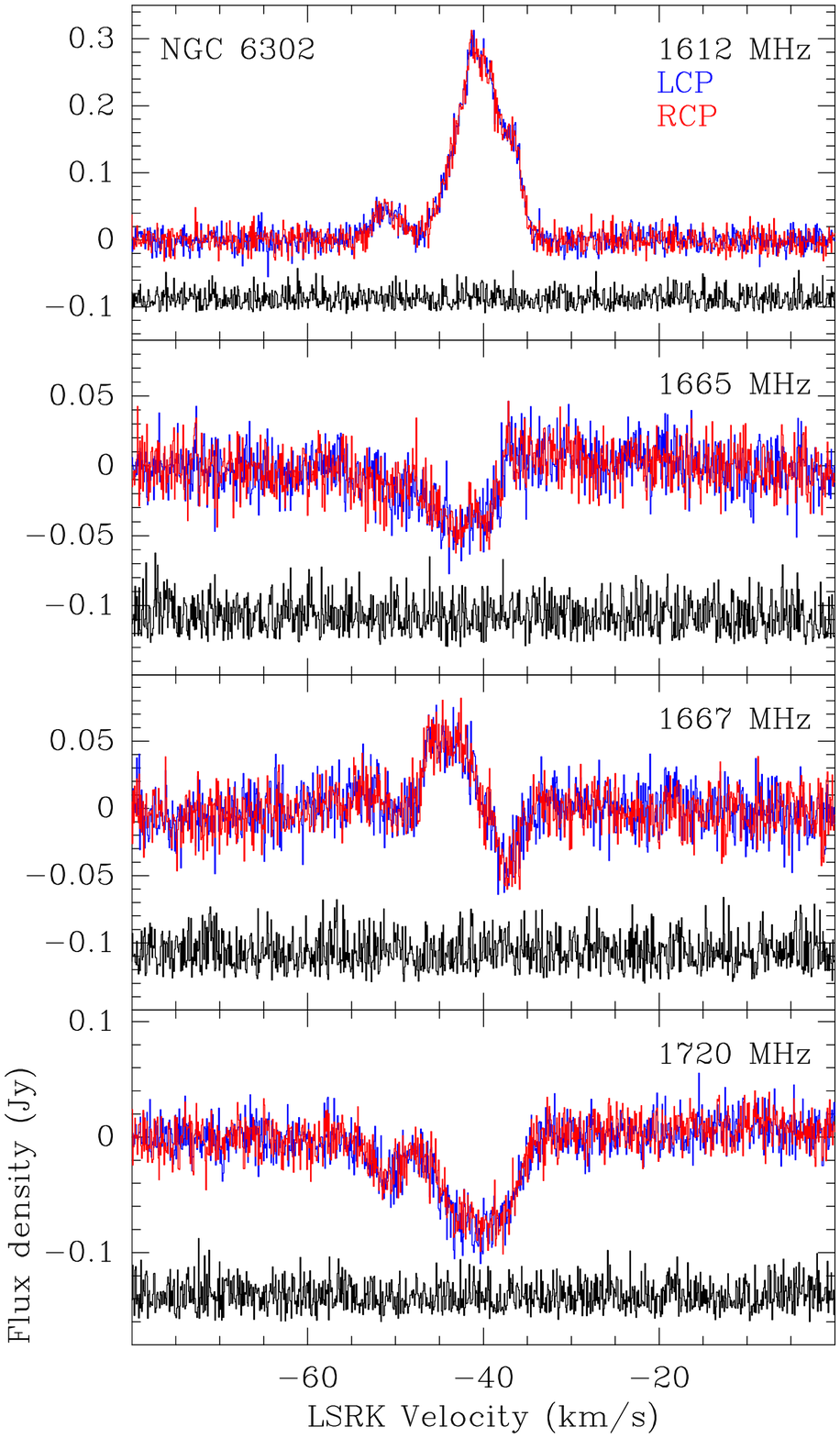}
	\caption{Left: OH spectra toward NGC 6302 in Stokes I and V. Right: Circular (red and blue) and linear (black) polarisations in NGC 6302. Linear polarisation has been arbitrarily displaced from a baseline level of 0 for better visualization.}
	\label{fig:ngc_spec}
\end{figure*}

Spectra of the four OH lines show similar profiles and flux density as in \cite*{pay88} and \cite{zij89}. The spectrum (Fig. \ref{fig:ngc_spec}) is seen in emission at 1612 MHz, in absorption at 1665 and 1720 MHz, and a mixture of emission and absorption at 1667 MHz. The spectrum at 1612 MHz shows two distinct components at $V_{\rm LSR}\simeq -51.4$ and $\simeq -41.5$ km s$^{-1}$, plus a possible additional component at $\simeq -36$ km s$^{-1}$, { blended with the one at $-41.5$ km s$^{-1}$}. In Table \ref{tab:ngc_1612} we list the parameters of the peak of the two clear components. These two components are also seen in CO spectra \citep{per07,din08}, with a third CO component (at $\simeq -22$ km s$^{-1}$) missing in OH. These CO components seem to trace an expanding disc (the missing CO component would be the receding hemisphere, not seen in OH probably due to continuum opacity).
The spectrum at 1720 MHz is similar, but in absorption, to the one at 1612 MHz. This similarity of the OH profiles at 1612 MHz with the one in absorption at 1720 MHz and that of CO suggests that the emission at 1612 MHz is thermal, and not maser. Moreover, the velocity gradient and position of the two main OH components (Fig. \ref{fig:ngc_cont}) are consistent with those of CO, so it is probably tracing the same structure. 

We do not see any polarisation in the OH lines. The same negative results were obtained by \cite{szy04} and \cite{wol12}, although our upper limits are more stringent by a factor of $\simeq 3$. 

%

\subsection{JaSt 23}

\subsubsection{Radio continuum}

Our radio continuum map shows a point source, with flux density $1.51\pm 0.06$ mJy. The flux density reported by \cite{van01} was $3.7\pm 0.4$ mJy at 6 cm, and $3.9\pm 0.4$ at 3 cm (source size $<1.80''$). Their reported position is compatible with ours. 

\subsubsection{OH lines}

\begin{figure*}
	\includegraphics[width=1.1\columnwidth]{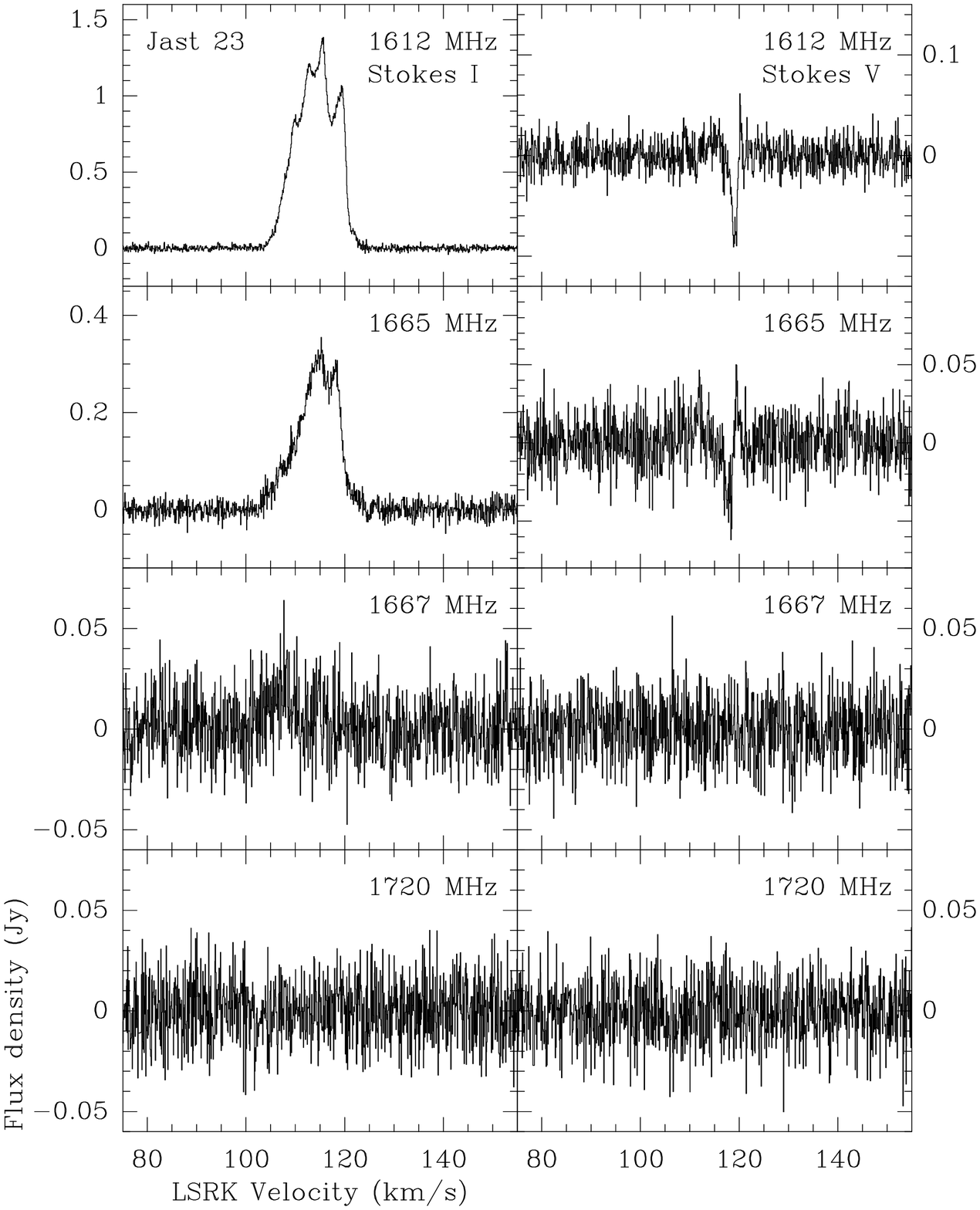}
	\includegraphics[width=0.8\columnwidth]{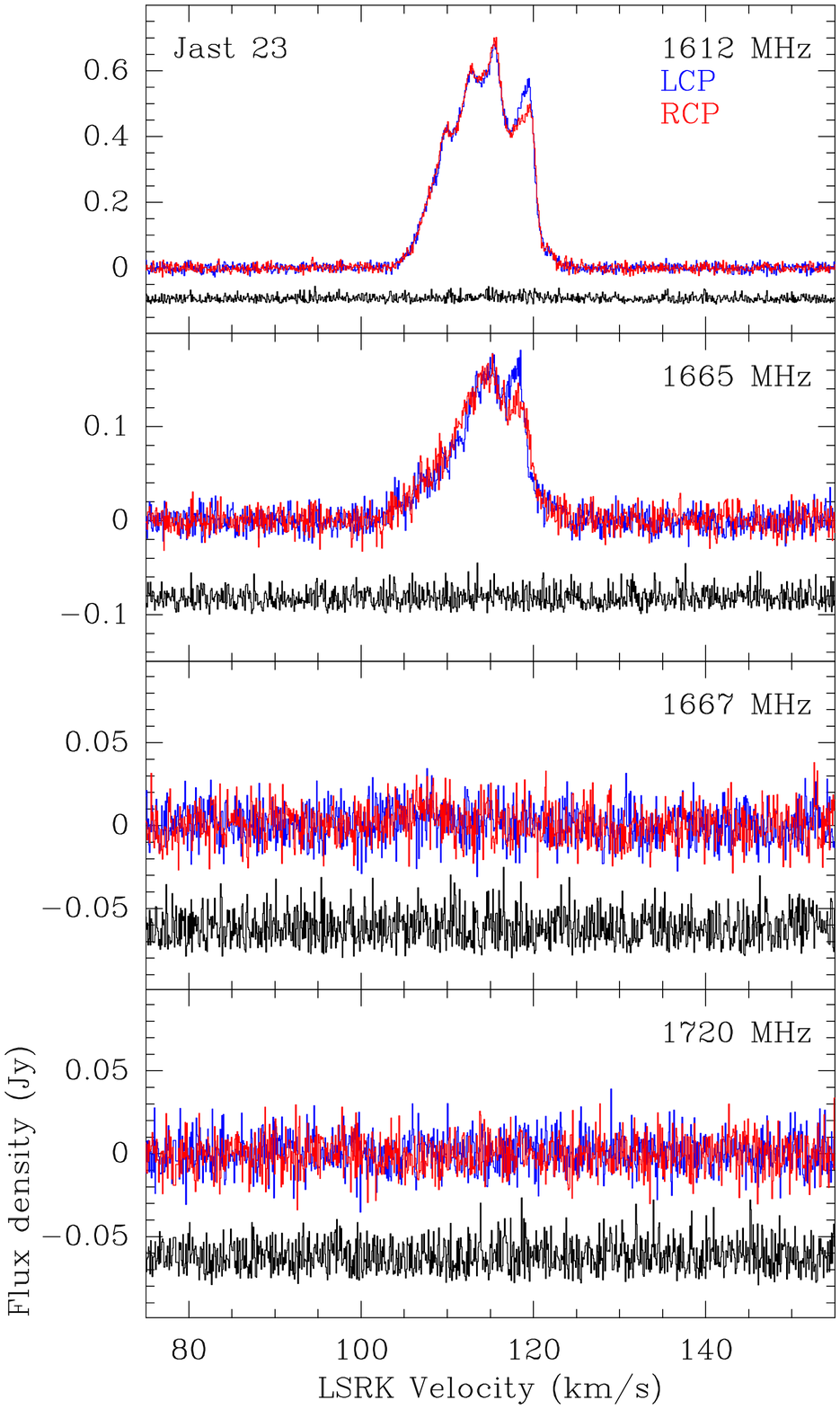}
	\caption{Left: OH maser spectra toward JaSt 23 in Stokes I and V. Right: Circular (red and blue) and linear (black) polarisations in JaSt 23. Linear polarisation has been arbitrarily displaced from a baseline level of 0 for better visualization.}
\label{fig:jast_spec}
\end{figure*}

We clearly detect OH emission at 1612 and 1665 MHz, with their spectra consisting in several velocity components (Fig. \ref{fig:jast_spec} and Table \ref{tab:jast_masers}). Weak emission at 1667 MHz seems also to be present, at $V_{\rm LSR}\simeq 107.2$ km s$^{-1}$.
\cite{sev97}, with a coarser spectral resolution, detected a single, wide spectrum at 1612 MHz (source OH 0.344 +1.567) centered at 115.2 km s$^{-1}$. 
\cite{gon14} did not detect polarised emission at either 1665 or 1667 MHz.

%

In our spectra, most of the detected OH emission is unpolarised, except for
the component at $\simeq 118-119$ km s$^{-1}$ in both the 1612 and 1665 MHz transitions,  which shows some circular polarisation (maximum polarisation fractions $p_c=0.08$ and $0.13$ at 1612 and 1665 MHz, respectively).
The spectra of the V Stokes parameter have low signal to noise ratio, but its shape at 1665 MHz might be due to Zeeman splitting. Assuming a Zeeman shift lower than the { Doppler linewidth}, a fit to equation \ref{eq:fit},  gives a value of $\simeq 0.8$ mG for the magnetic field, for saturated maser emission and planar morphology. If the shift of the circularly polarised emission where larger than the Doppler linewidth, the estimated magnetic field would be  $\simeq 3$ mG. 

{ Both solutions (Zeeman shift larger or smaller than the Doppler linewidth) are, in principle, possible. OH maser emission arises from the outer, cold layers of circumstellar envelopes \citep[a few times $10^{16}$ cm from the central star,][]{eli82}, with gas kinetic temperatures $\simeq 50-150$ K \citep{eli78,ces91}, corresponding to thermal OH linewidths of $\simeq 0.4-0.6$ km s$^{-1}$. This is larger than the shift of $\simeq 0.24$ km s$^{-1}$ in each of the circularly polarised components that would be created by a magnetic field strength of $\simeq 0.8$ mG, but smaller than the shift of $\simeq 0.9$ km s$^{-1}$ for the case of $B=3$ mG.}

\subsection{IRAS 17375-2759}

This source was not one of our main targets, but it is
located $12.3'$ from the phase centre of the observation toward JaSt 23, i.e., within the primary beam of ATCA (FWHM $\simeq 30'$ at 1612 MHz). Unlike the other five sources observed in this paper, which are confirmed OHPNe \citep{usc12}, the nature of IRAS 17375-2759 is uncertain. \cite{usc12} list it as a possible OHPN, since it presents both radio continuum and OH emission, but there is no confirmation that the source is a PN.

\subsubsection{Continuum}

We detect unresolved radio continuum emission, with a flux density of $4.78\pm 0.15$ mJy (obtained after applying the correction of the primary beam response). \cite{pot88} reported a point source of 2.6 mJy at 6 cm. Our higher flux at longer wavelength is not expected if the emission is due to free-free processes in an ionized nebula, unless the emission is variable. Possible explanations of the higher flux density found in our more recent observations could be the expansion of a dense ionized nebula { \citep[e.g.][]{kwo81,cri98,gom05}} or the presence of non-thermal emission processes { \citep[e.g.][]{per13,sua15}}. { Quasiperiodic variability due to changes in mass-loss processes is also possible \citep{san04,cer11}.} Simultaneous observations at different wavelengths would be needed to ascertain the nature of the emission.

\subsubsection{OH lines}

\begin{figure*}
	\includegraphics[width=1.1\columnwidth]{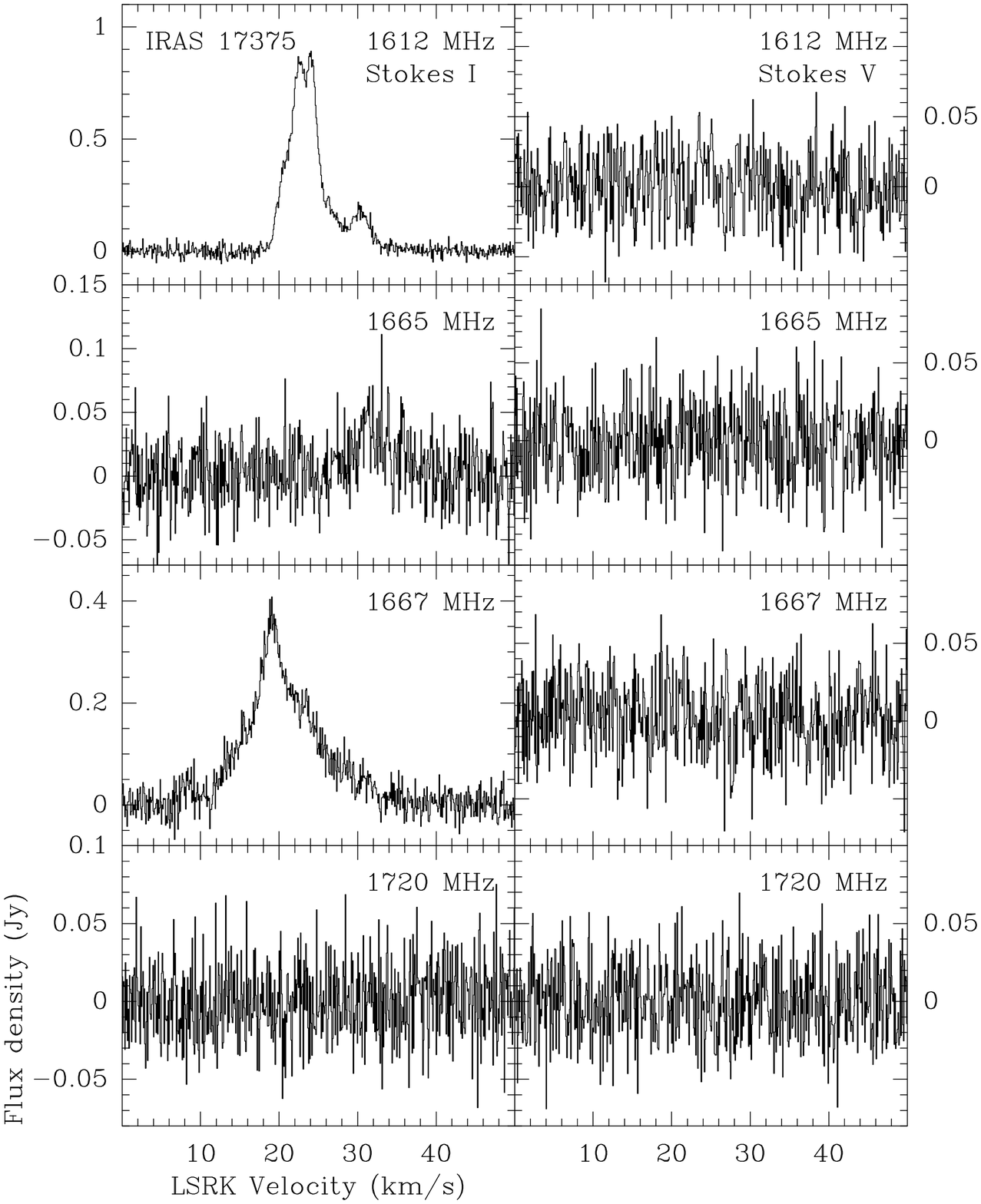}
	\includegraphics[width=0.8\columnwidth]{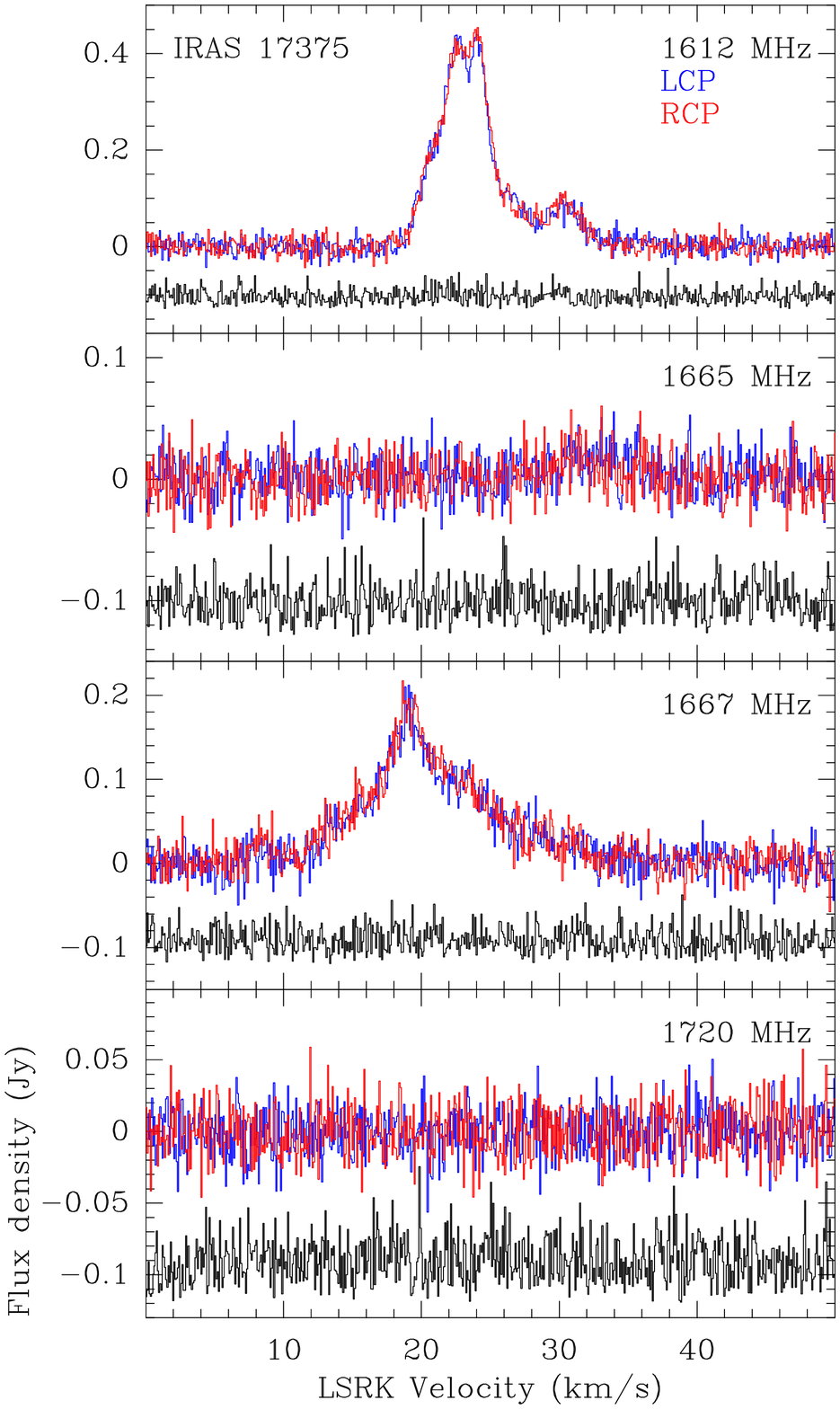}
	\caption{Left: OH maser spectra toward IRAS 17375-2759 in Stokes I and V. Right: Circular (red and ble) and linear (black) polarisations in IRAS 17375-2759. Linear polarisation has been arbitrarily displaced from a baseline level of 0 for better visualization.}
	\label{fig:iras0_spec}
\end{figure*}

We detect OH emission at 1612 and 1667 MHz (Fig. \ref{fig:iras0_spec}). Three and two distinct components are identified in the 1612 and 1667 MHz spectra, respetively (Table \ref{tab:iras0_masers}). Our spectra are very similar to those reported by \cite{szy04} and \cite{wol12} using the Nan\c{c}ay antenna. However, unlike these authors, who report circular polarisation fractions of $\simeq 7-15$\% at 1612 MHz,  we do not see any significant polarisation at either 1612 or 1667 MHz, with upper limits below these reported values.

%


\subsection{IRAS 17393-2727 (OH 0.9+1.3)}
\subsubsection{Continuum}

\begin{figure}
	\includegraphics[width=0.9\columnwidth]{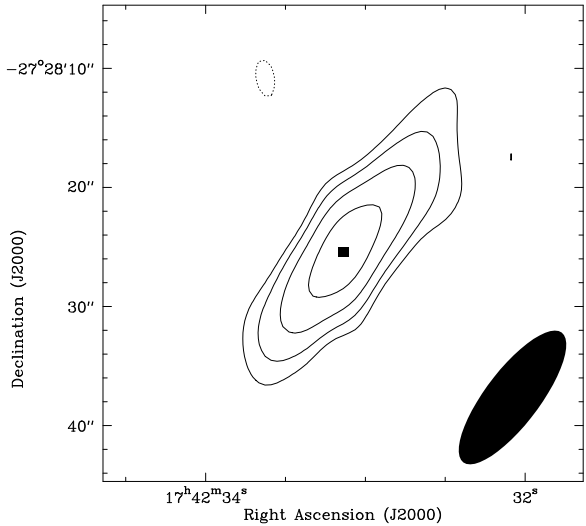}\\
			\caption{Contour map of the continuum emission in IRAS 17393-2727. The contour levels are $3\times 2^n$ times 25 $\mu$Jy beam$^{-1}$ (the rms of the map), with $n=0$ to 3. The black square  marks the  location of peak emission at 1612 MHz. The other maser components appear at the same location in the image. The filled ellipse represents the synthesized beam of the continuum image ($13.6''\times 4.5''$, p.a. $=-37^\circ$).}
	\label{fig:iras_cont}
\end{figure}

We detected a radio continuum source with 
flux density $0.86\pm 0.10$ mJy (Fig. \ref{fig:iras_cont} and Table \ref{tab:cont}), which is consistent with other data in the literature \citep*[$<3$ mJy at 2 cm, 1.4 mJy at 6 cm,][]{pot87}, for an ionized region. The main NW-SE elongation seen in Fig. \ref{fig:iras_cont} is obviously due to the beam elongation. The deconvolved size of the source is  $2.9''\times 1.6''$, p.a. $=-26^\circ$, with an orientation close to the optical nebula seen in HST images \citep{man11}. In addition, there is a bulging in the lowest contours, along p.a. $=13^\circ$, which may correspond to a weaker extension of the ionized nebula.

\subsubsection{OH lines}

\begin{figure*}
	\includegraphics[width=1.1\columnwidth]{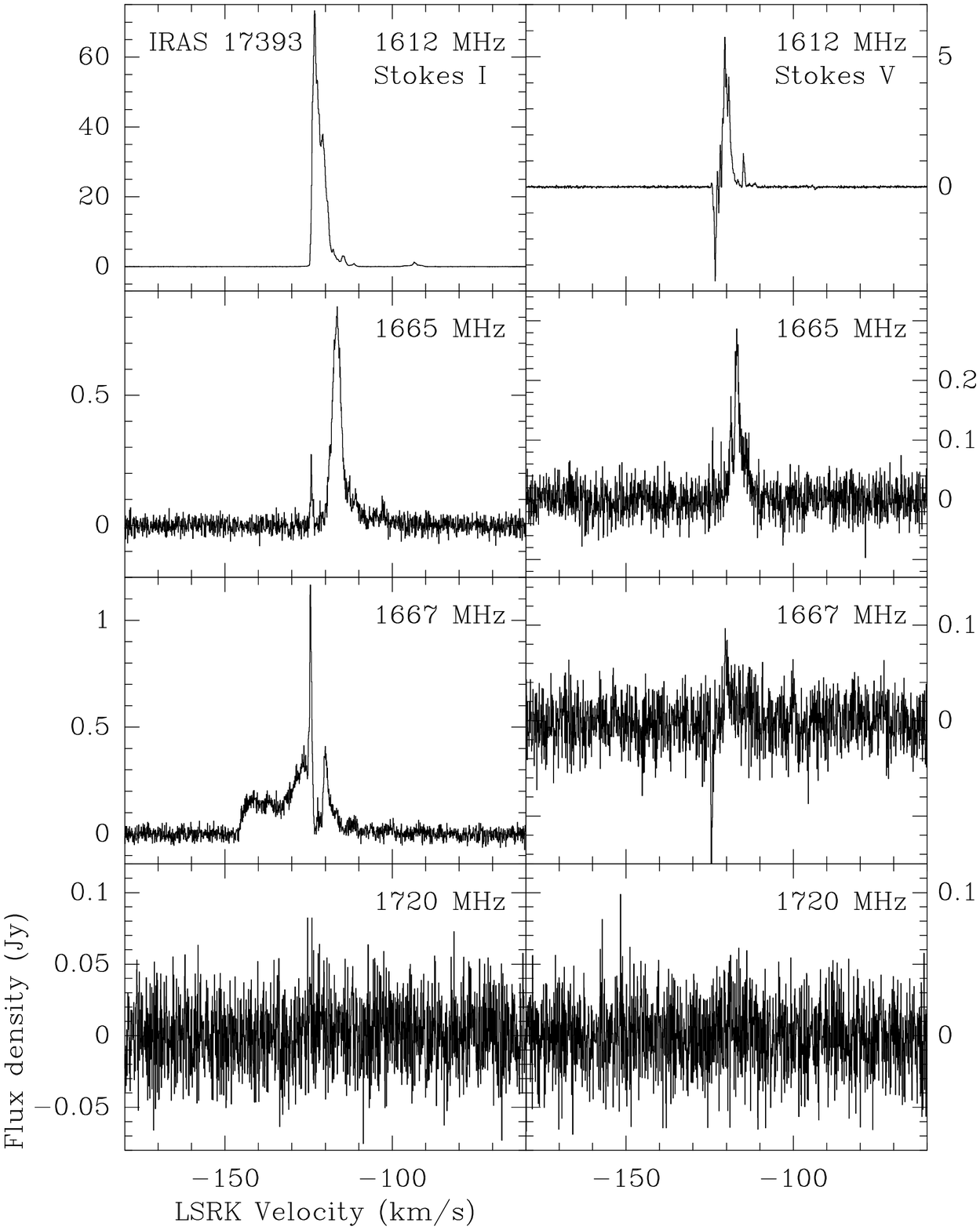}
	\includegraphics[width=0.8\columnwidth]{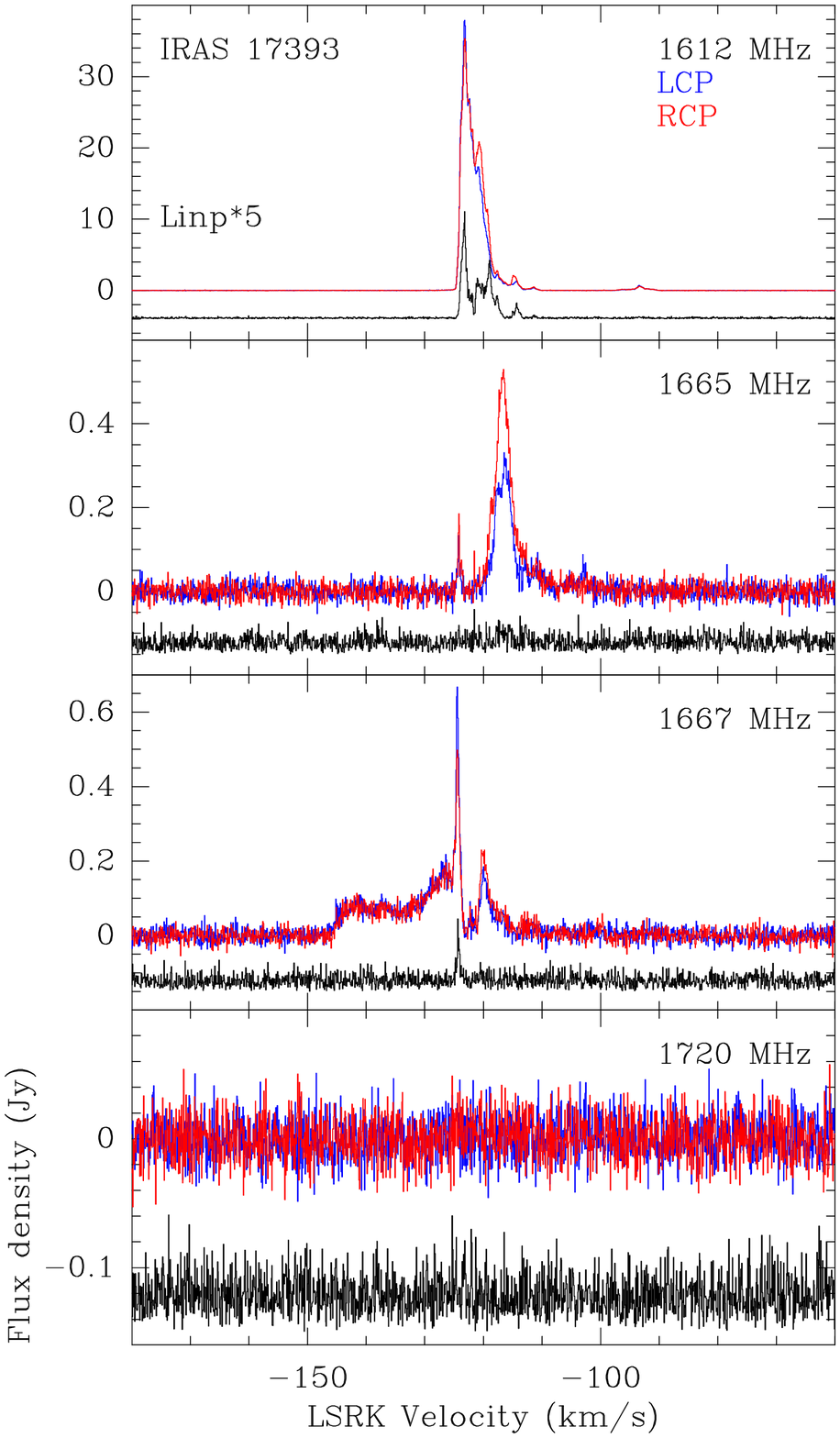}
	\caption{Left: OH maser spectra toward IRAS 17393-2727 in Stokes I and V. Right: Circular (red and blue) and linear (black) polarisations in IRAS 17393-2727. Linear polarisation has been multiplied by 5, and arbitrarily displaced from a baseline level of 0, for better visualization.}

	\label{fig:iras_spec}
\end{figure*}

We detected multiple components at 1612, 1665, and 1667 MHz (Fig. \ref{fig:iras_spec}). The spectrum at 1612 MHz is variable, and we find the highest peak flux density reported so far in the literature \citep{pot87,tel89,ker74,mig95,sev97,wol12}. In fact, it shows a steady increase, of about 1 Jy per year \citep{mig95}. We note a wide plateau of emission at 1667 MHz, of $\simeq 0.14$ Jy at $V_{\rm LSR} \simeq -145$ to -130 km s$^{-1}$. \cite{zij89} also mention the presence of a weak plateau in the the spectrum at 1667 MHz, over the velocity range covered by the emission at 1612 MHz. The plateau we detected at 1667 MHz is stronger and extends beyond the velocity range of OH at 1612 MHz.


%

\begin{figure}
	\includegraphics[width=0.8\columnwidth]{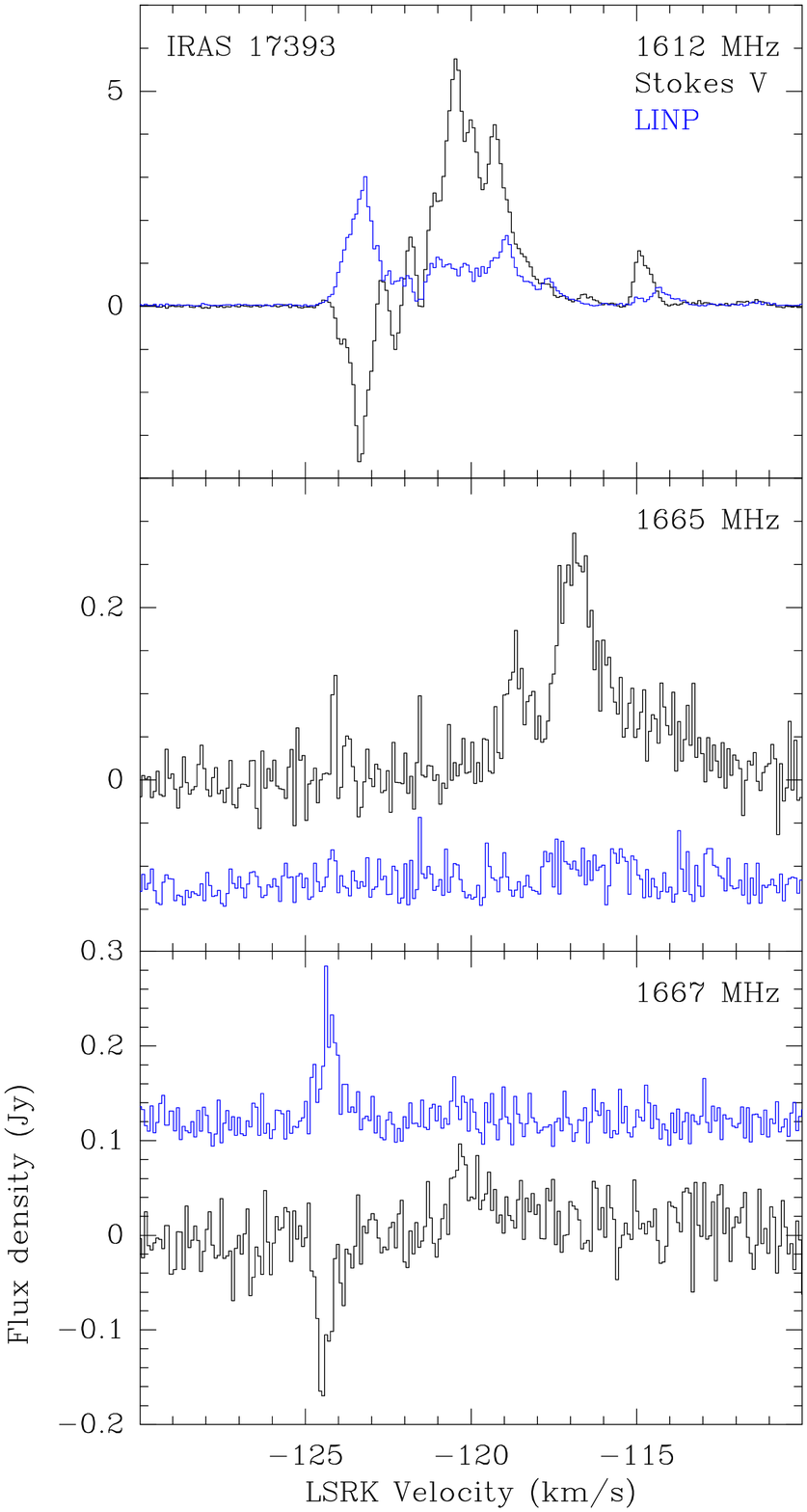}
	\caption{Spectral zoom of circular (Stokes V, black line)  and linear polarisation (blue line) toward IRAS 17393-2727, at 1612, 1665, and 1667 MHz. In the two last panels, linear polarisation has been arbitrarily displaced from a baseline level of 0 for better visualization.}
	\label{fig:iras_spec_zoom}
\end{figure}

We detect circular polarisation at all three transitions, and significant linear polarisation at 1612  and 1667 MHz (Fig. \ref{fig:iras_spec}, \ref{fig:iras_spec_zoom}). The polarisation levels are similar to those found in the past with single-dish antennas \citep{szy04,wol12}. The spectra of Stokes V at both 1612 and 1667 MHz shows S-shaped spectra suggestive of Zeeman splitting. No reasonable fit was found assuming that the Zeeman shift is smaller than the Doppler linewidth. If we assume it is larger, we estimate a magnetic field of 12-24 mG for the emission at $-120$ to $-123$ km s$^{-1}$, and 6 mG for that at $\simeq -93$ km s$^{-1}$.

{ The position angles of the linear polarisation (Table \ref{tab:iras_masers})   do not show any obvious trend from which to compare the magnetic field orientation with the morphology of the nebula. The scatter among the different OH components is so large that it seems random. The absence of any ordered pattern in the magnetic field direction may be due to a local mass-loss outburst affecting the OH-emitting region \citep{wol12}.}

\subsection{Vy 2-2}
\subsubsection{Continuum}

We detected an unresolved radio continuum source, with a flux density of $16.53 \pm 0.24$ mJy. The radio continuum emission from this source has been extensively studied in the literature. The flux density we found is consistent with that reported elsewhere at similar wavelengths \citep[e.g.,][]{pur82,sea83,sea91,con98}. \cite{cri98} found variability in its flux density, specially at $\simeq 1.5$ GHz, which can be explained as expansion of the ionized nebula.

\subsubsection{OH lines}

\begin{figure*}
	\includegraphics[width=1.1\columnwidth]{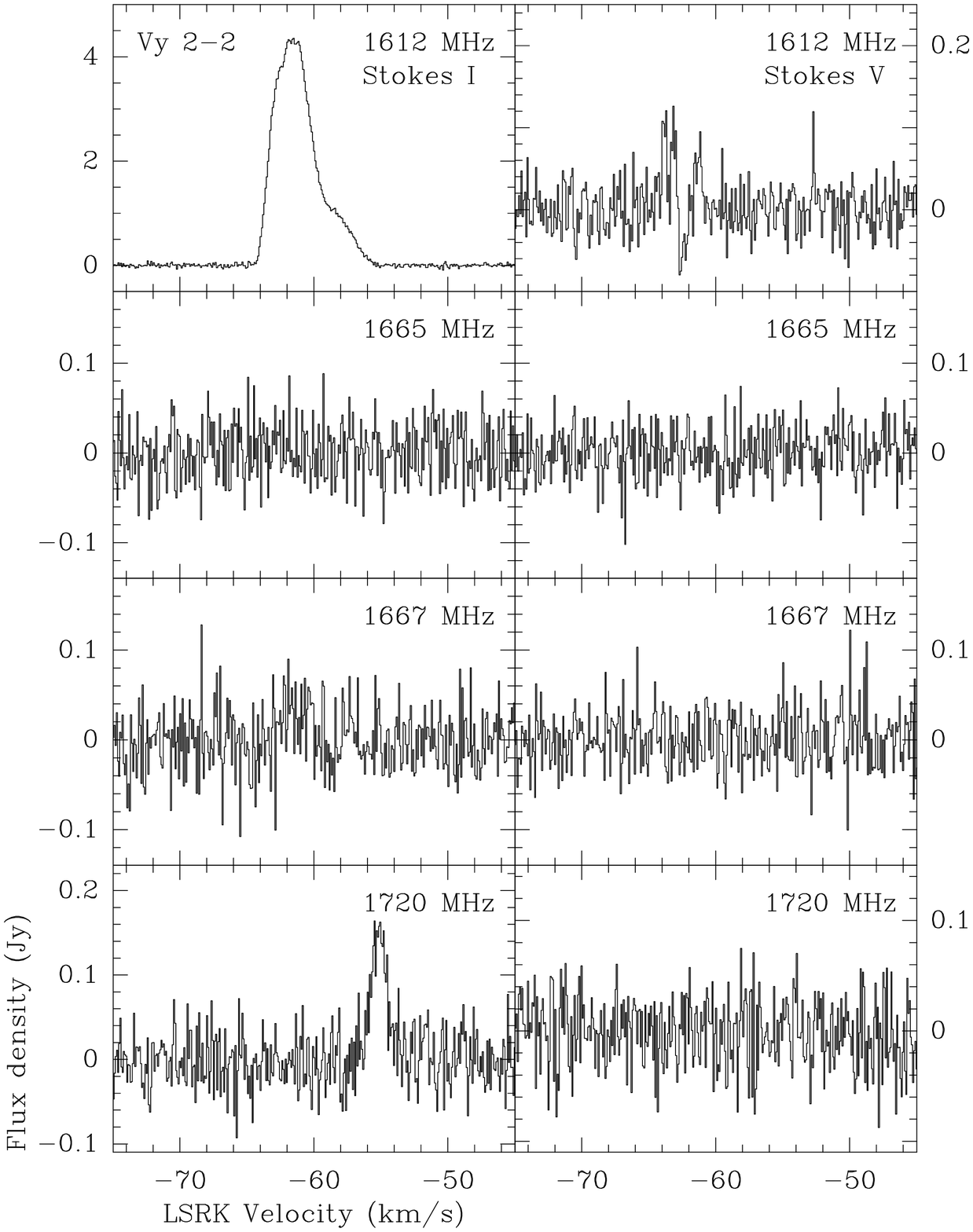}
	\includegraphics[width=0.815\columnwidth]{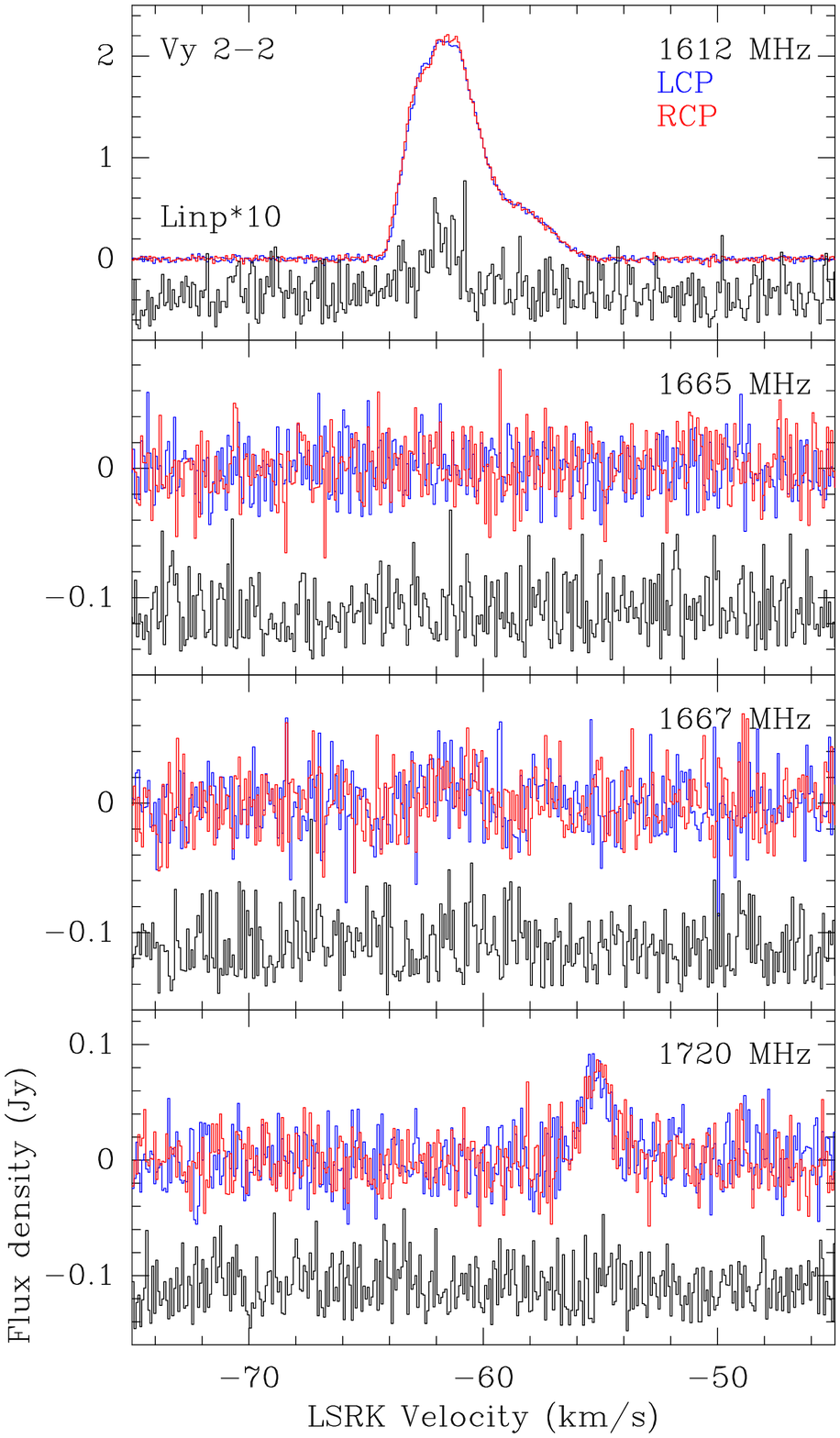}
	\caption{Left: OH maser spectra toward Vy 2-2 in Stokes I and V. Right: Circular (red and blue) and linear (black) polarisations in Vy 2-2. Linear polarisation has been multiplied by 10 and arbitrarily displaced from a baseline level of 0 for better visualization.}
	\label{fig:vy_spec}
\end{figure*}

This source shows a wide profile  of OH emission at 1612 MHz (extending for a range of $>9$ km s$^{-1}$, Fig. \ref{fig:vy_spec} and Table \ref{tab:vy_masers}), with a similar flux density as reported before \citep[e.g.,][]{dav79,sea83,zij89}. Moreover, we clearly detect a single component of OH emission at 1720 MHz. This type of emission is extremely rare in evolved stars, and so far only two other PNe, K 3-35  \citep[][see also section \ref{sec:k3maser} in this paper]{gom09} and IRAS 16333-4807 \citep{qia16}, have been reported to show emission at this transition. We do not detect emission at 1665 or 1667 MHz.

The emission at 1612 MHz seems to be polarised, with levels $0.013$ and  $0.024$ (circular and linear polarisation, respectively). The level of linear polarisation is similar to that found by \cite{szy04} and \cite{wol12}, but the circular polarisation is lower in our data, by a factor of $\simeq 5-8$.

%

\subsection{K 3-35}

\subsubsection{Continuum}

\begin{figure}
	\includegraphics[width=0.9\columnwidth]{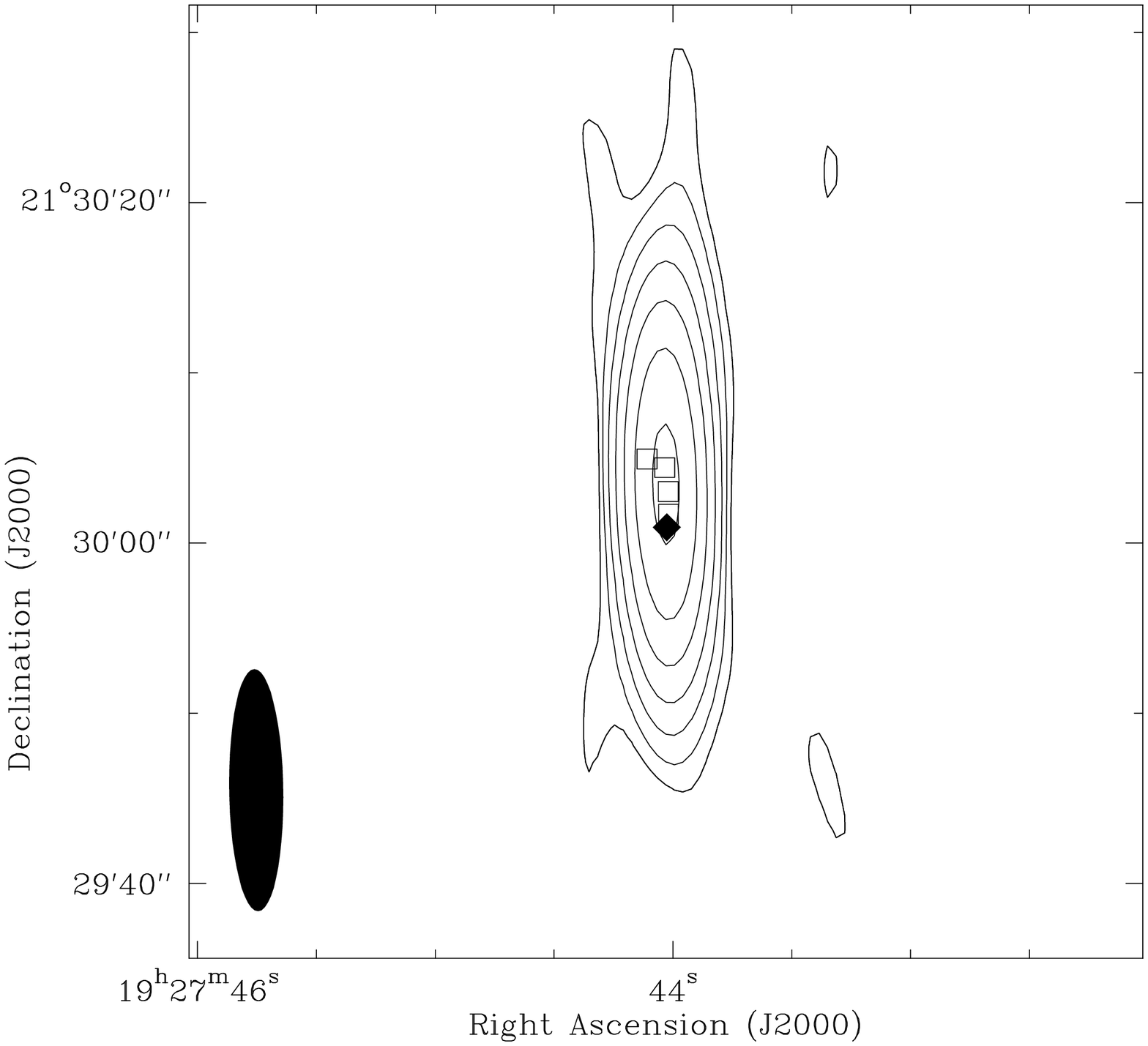}\\
	\caption{Contour map of the continuum emission in K 3-35. The contour levels are $3\times 2^n$ times 50 $\mu$Jy beam$^{-1}$ (the rms of the map), with $n=0$ to 6. The open squares mark the location of the OH peaks at 1612 MHz. The filled diamond marks the peak at 1720 MHz. The filled ellipse represents the synthesized beam of the continuum image ($14.2''\times 3.2''$, p.a. $=1^\circ$)}
	\label{fig:k3_cont}
\end{figure}

We found an unresolved continuum source with flux density $11.9\pm 0.3$ mJy (Fig. \ref{fig:k3_cont}). A gaussian fit gives a deconvolved size of $ 2.4\times 0.5''$, p.a. $=7^\circ$ for this source. 
The radio continuum emission from this source has been extensively studied in the past, but the observations closest in frequency to ours are those of \cite{aaq93}, \cite{con98} and
\cite{gom09}, with flux densities of $14.6\pm 0.6$ (1.4 GHz), 20 mJy (1.5 GHz) and $18\pm 1$ mJy (1.7 GHz), as well as deconvolved sizes of $1.96''\times 0.48''$, p.a. $=13^\circ$ (1.5 GHz) and
$\simeq 1.8''\times 0.5''$, p.a. $=16^\circ$ (1.7 GHz). Given the different central frequency and bandwidths of the observations, and the calibration uncertainties, we can not conclude that this source has changed significantly.

\subsubsection{OH lines}

\label{sec:k3maser}
\begin{figure*}
	\includegraphics[width=1.1\columnwidth]{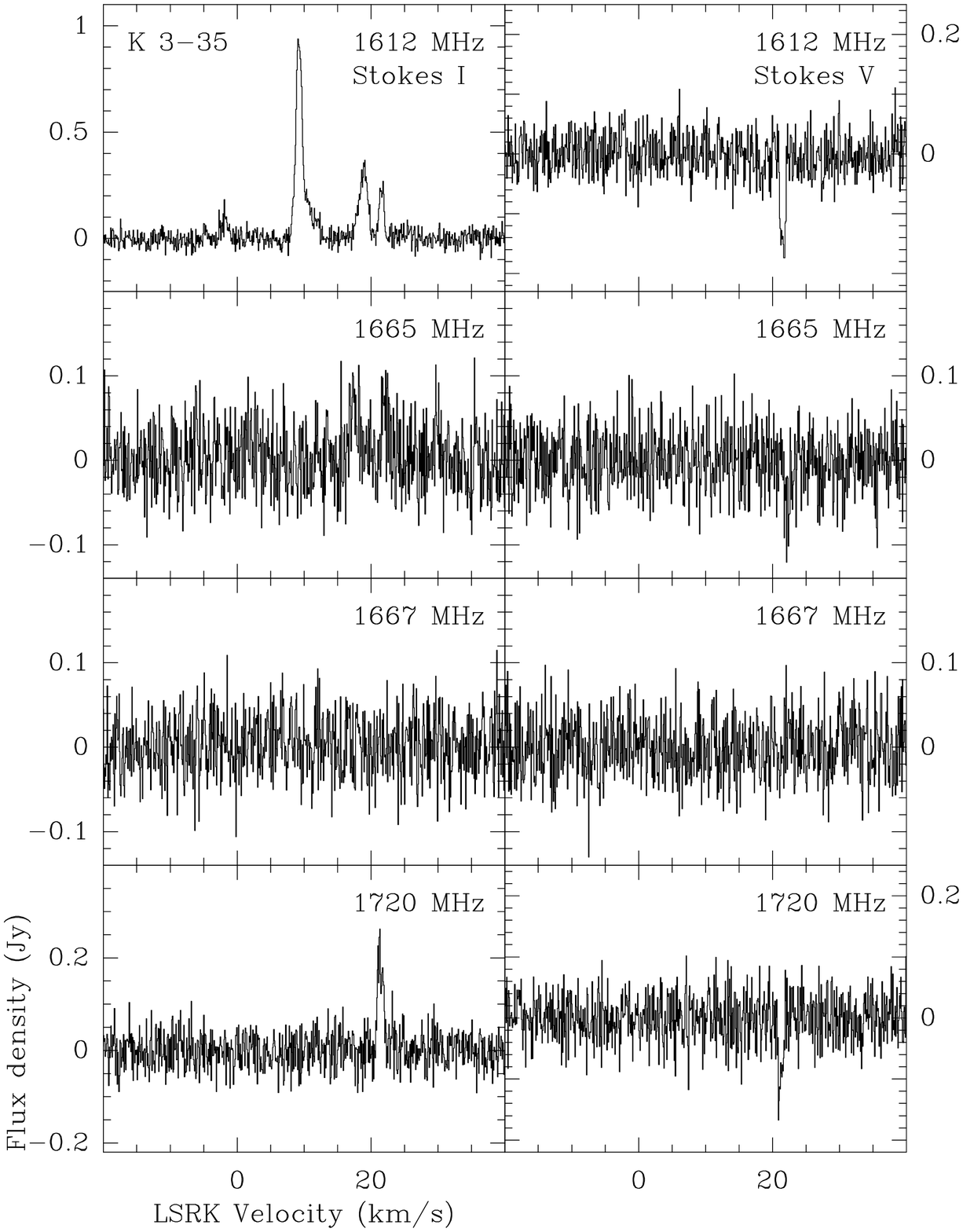}
	\includegraphics[width=0.82\columnwidth]{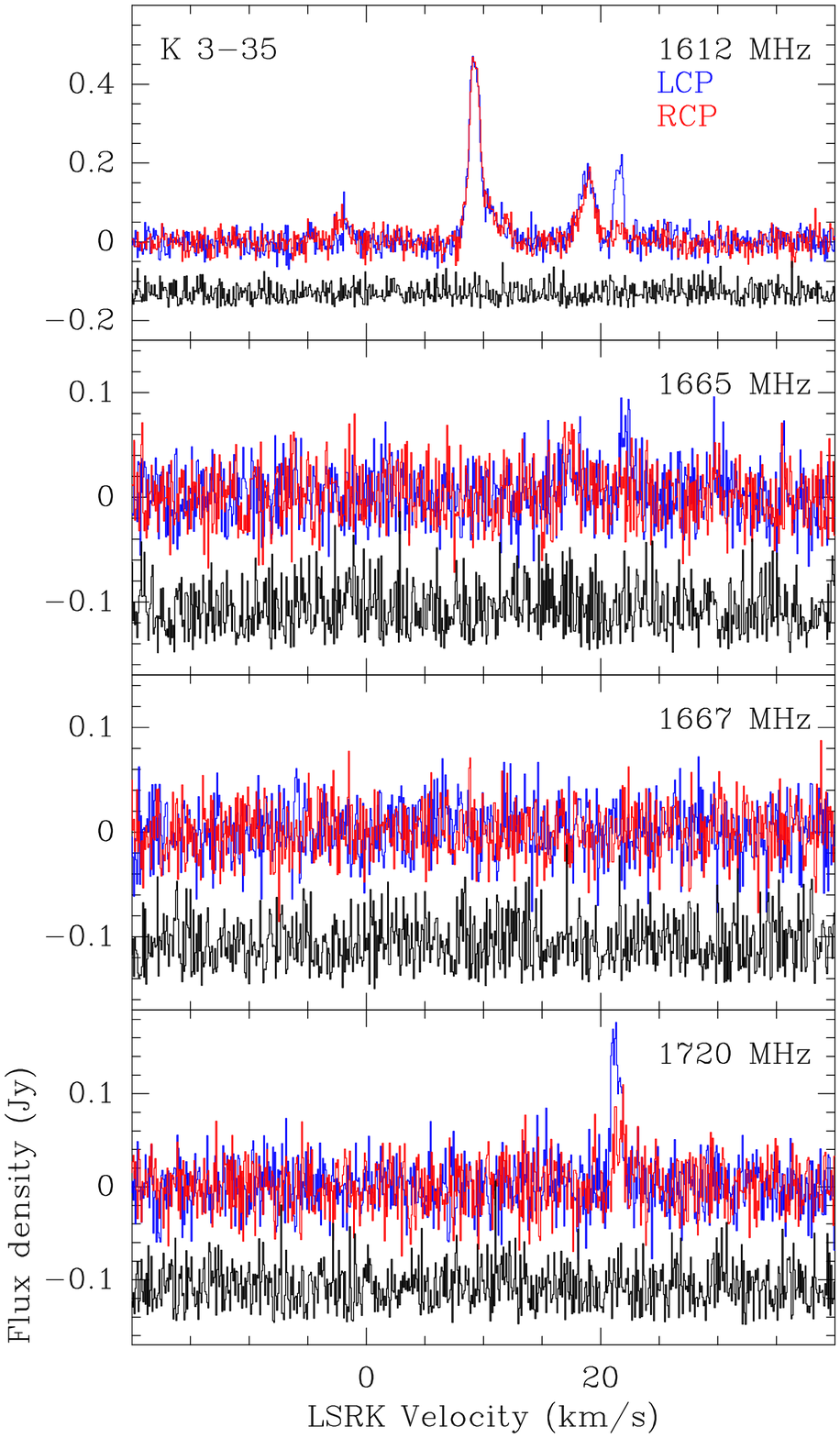}
	\caption{Left: OH maser spectra toward K 3-35 in Stokes I and V. Right: Circular (red and blue) and linear (black) polarisations in K 3-35. Linear polarisation has been arbitrarily displaced from a baseline level of 0 for better visualization.}
	\label{fig:k3_spec}
\end{figure*}

The shape of the spectra of Stokes I of all four transitions (Fig. \ref{fig:k3_spec} and Table \ref{tab:k3_masers}) are similar to those reported in previous works \citep[e.g.,][]{gom09,des10}, although the flux density is significantly lower in all cases. The position of all our maser components are compatible among them and with the continuum peak, within the errors.

We only detect significant polarisation in the components around $V_{\rm LSR}\simeq 21-22$ km s$^{-1}$ at 1612, 1665, and 1720 MHz. Its polarisation fraction is $0.71\pm 0.25$ and  $0.50\pm 0.23$ at 1612 and 1720 MHz, respectively. The signal-to-noise ratio of the signal in Stokes V is too low to get a meaningful estimate of this polarisation fraction at 1665 MHz.

We notice some differences in the polarisation properties we see and those reported by \cite{gom09}. First, the sign of the Stokes V signal at 1720 MHz is reversed: in our data, the 1720 MHz emission is dominated by LCP, but it was by RCP in \cite{gom09}. We cannot confirm the Zeeman splitting suggested by these authors in the 1665 MHz transition, although the significant decrease of flux density (a factor of 4) in our data makes this spectrum very noisy. Overall, we see no sign of Zeeman splitting in any of the OH transitions.

\section{Discussion}

With the observations in this paper, we now have interferometric polarimetric observations of OH emission in all known OHPNe: those presented in our paper (except IRAS 17375-2735, which is not a confirmed member of this class), together with IRAS 17347-3139 \citep{taf09} and IRAS 16333-4807 \citep{qia16}. Therefore, we can know discuss some interesting properties of these sources as a group.

\subsection{Spectral shapes and evolution}

As pointed out by other authors \citep[e.g.,][]{zij89} all OHPNe show spectral shapes that differ from the double-peaked profile typical of AGB stars. While those double-peaked profiles are believed to arise from the approaching and receding sides of the AGB envelope \citep{rei76}, the departure from that spectral pattern in post-AGB stars and PNe is interpreted as evidence of the presence of non-spherical mass-loss \citep*{dea04}.

Moreover, we note significant differences also among the spectral profiles of OHPNe, specially at 1612, 1665, and 1667 MHz. The OH spectrum of some sources is dominated by separate, narrow features ($\Delta V\la 1$ km s$^{-1}$), easily identifiable in the spectrum. This would be the case of K 3-35 (Fig. \ref{fig:k3_spec}), IRAS 17347-3139 \citep{taf09}, and IRAS 16333-4807 \citep{qia16}. On the other hand, the rest of the confirmed OHPNe (NGC 6302, JaSt 23, IRAS 17393-2727, Vy 2-2) show smoother and wider spectra. These smoother spectra could be the result of blending of multiple narrow features, spectrally and spatially unresolved, as suggested by the presence of local maxima across the spectra. If these features are numerous and with small velocity separations, they could give rise to complex spectra, as opposed to a few, well-separated features.

We note that all the OHPNe with narrow-feature spectra are also emitters of water masers \citep{mir01, deg04,usc14}. A possible exception is IRAS 17393-2727, which might also harbour water masers \citep{gom15}, but shows complex OH spectra. The association of water masers with this source has not been yet confirmed with interferometer observations. 

These PNe showing both OH and water masers has been suggested to be the youngest among the group of maser-emitting PNe \citep{gom15}.
Therefore, it seems that OH emission at the beginning of the PN phase is characterized by narrow, well-defined spectral features. Considering that OH masers are pumped under very particular physical  conditions, it means that the regions of the circumstellar envelope where these masers are pumped in the youngest OHPNe are relatively limited. As the PNe evolve, larger regions of the envelope can show OH emission, then producing more complex spectra. It is even possible that there is significant contribution of thermal OH emission from dense regions, in addition to OH masers. 

Specially in the case of NGC 6302, the OH emission could be dominated by (or being entirely) of thermal emission. This questions its inclusion within the same class of sources as other PNe clearly showing maser emission, with narrow and variable spectral features (e.g., K 3-35, IRAS 16333-4807, IRAS 17347-3139). This would explain the apparent discrepancy between the supposed extreme youth of PNe with OH emission, and the large size and optical brighness of NGC 6302, which suggested that it was in a more evolved stage than other OHPNe. Later in the PNe evolution, the growth of dense circumstellar toroids could give rise to detectable thermal OH emission (as in NGC 6302), but these objects should not be confused with the younger OH-maser emitters, as may not have much in common from a morphological or evolutionary point of view.

We note that OH emission at 1720 MHz shows narrow features in all sources where it has been detected. This type of emission is discussed in section \ref{sec:1720}.

\subsection{Polarisation and evolution}

We detected circular polarisation in all confirmed OHPN sources except NGC 6302. The unconfirmed OHPN IRAS 17375-2735 does not show any polarisation. Significant linear polarisation is only detected in IRAS 17393-2727 and Vy2-2. OHPNe show, in general, low levels ($\la 25$\%) of circular polarisation. Higher polarisation fractions are only seen in IRAS 17393-2727 (up to 36\%), K 3-35 (72\%) and IRAS 16333-4807 \citep[$\simeq 100$\%,][]{qia16}. We do not see a clear pattern with regards to these polarisation fractions. There seems to be a trend of higher circular polarisation fractions in sources that show water masers in addition to OH emission, but IRAS 17347-3139 is a outstanding exception, since it does not show any significant circular polarisation \citep{taf09}.

\cite{wol12} found that post-AGB stars and PN have highest fractional polarisation, compared with AGB (OH/IR) stars. Here we found a possible trend that younger PNe have higher polarisation fractions than the more evolved ones. 
{ However, the lack of confirmed Zeeman pairs in OHPNe \citep[only those in IRAS 16333-4807 seem indisputable,][]{qia16} precludes a quantification of magnetic field strengths to support whether this apparent trend in fractional polarisation may indicate a maximum in the magnetic field strength around the end of the post-AGB and early PN phase}. Interferometric observations of OH with Very Large Baseline Interferometry would be necessary to spatiality resolve the polarised emission, to isolate Zeeman pairs.

\subsection{OH emission at 1720 MHz in PNe}

\label{sec:1720}

Of particular interest is the detection of OH emission at 1720 MHz in Vy 2-2. This transition is commonly seen associated with shocks created by supernova remnants on molecular clouds \citep{fra96,gre97} and in star-forming regions \citep[e.g.,][]{wal16}. However, it is extremely rare in low and intermediate-mass evolved stars. No AGB star has been reported to show emission at 1720 MHz \citep{sev01}, and it has been confirmed in only one post-AGB star \citep[OH 9.1-0.4][]{sev01,dea04}. On the other hand, this paper completes the census of 1720 MHz emission in all OHPNe sources, which has yielded its detection in three sources (out of the seven confirmed OHPNe): K 3-35 \citep[][and Fig.\ \ref{fig:k3_spec}]{gom09}, IRAS 16333-4807 \citep{qia16}, and Vy 2-2 (Fig.\ \ref{fig:vy_spec}). 

Thus, it is clear that considering evolved sources showing OH emission at the other ground-state transitions, the emission at 1720 MHz is far more common in PNe than in AGB and post-AGB stars. This transition is thought to be collisionally excited in C-shocks \citep*{loc99,war12}, which are non-dissociative shocks produced in the presence of magnetic fields \citep{dra80}. This is in contrast with the other three ground-state OH transitions, which are radiatively pumped \citep{eli76,col94}. 
The particular incidence of emission at 1720 MHz in PNe could be related to particular mass-loss processes ocurring after the beginning of photoionization.
\cite{qia16} suggested that the presence of OH emission at 1720 MHz could trace the ejection of equatorial mass-loss, during the early PN phase. 

Considering that the detected spectrum of Vy 2-2 at 1720 MHz has only a component, and it is located very close to the centre of the radio continuum emission, it is not possible from our data to determine whether it is actually arising from an equatorial ejection. Further observations with higher angular resolution would be necessary to determine the structure where this emission arises from. 

In the comparison of Vy 2-2 with the other two PNe with OH emission at 1720 MHz, a remarkable difference is that the former does not show any significant circular polarisation, whereas the polarisation fraction of this emission is $\simeq 50$\% in K 3-35 (Table \ref{tab:k3_masers}) and $\simeq 100$\% in IRAS 16333-4807 \citep{qia16}. If the emission at 1720 MHz is produced in magnetized shocks, one would expect it to be circularly polarised. It is possible that in Vy 2-2 the magnetic field in the shock is aligned near the plane of the sky.

{ Another interesting comparison can be made with respect to the excited-state OH emission at 6035 MHz. This emission has only been observed in two PNe:  Vy 2-2 \citep{jew85,des02} and K 3-35 \citep{des10}. Since these objects are also emitters of OH at 1720 MHz (note that there is no reported observation of excited-state OH towards IRAS 16333-4807), one can speculate whether the emission of these two transitions can be produced in the same regions and/or share a similar pumping mechanism. However, the behaviour of these two lines is different in these objects. While the spectra at both 1720 and 6035 MHz peak at a similar velocity in  K 3-35 \citep{des10}, suggesting that they might arise from the same region, there is a significant shift of $\simeq 8$ km s$^{-1}$ between the two transitions in Vy 2-2 \citep[][and Fig. \ref{fig:vy_spec}]{des02}. It is obviously impossible to draw firm conclusions from only two sources, but our results show that the OH emission at 1720 and 6035 MHz is not necessarily related. Sensitive studies of these two relatively little-studied transitions in evolved objects would be necessary to understand their excitation mechanism and locations.}

\section{Conclusions}

We have carried out interferometric, full-polarisation observations of the four ground-state transitions of OH at 18 cm, toward five OHPNe, plus one OHPNe candidate. Our conclusions are as follow:
\begin{enumerate}
\item We detect both radio continuum and OH emission at 1612 MHz in all sources. Emission or absorption at 1665 and 1667 MHz is detected in 3 and 2 sources, respectively. Emission at 1720 GHz is detected in Vy 2-2 (new detection) and K 3-35. Thus, Vy 2-2 is the third PNe known to harbour emission at this OH transition.
\item Circular polarisation is found in four of the six sources, while linear polarisation seems to be present in two of them. The polarisation fractions of these sources are, in general, low ($<25$\%), except for IRAS 17393-2727 and K 3-35.
\item Possible Zeeman pairs are found in JaSt 23 and  IRAS 17393-2727, yielding estimates of the magnetic fields strength along the line of sight of 0.8-3 and 6-24 mG, respectively.
\item { Our data suggest some evolutionary trends, with narrow maser features and higher polarisation fractions in the younger PNe. OHPN sources which are somewhat more evolved show wider, more complex spectra, probably with a significant contribution from thermal OH emission. }
\item Shock-excited OH emission at 1720 MHz seems to be more common in PNe than in AGB and post-AGB stars. This could be related to the presence of equatorial ejections during the early PN phase.
\end{enumerate}

\section*{Acknowledgements}

The Australia Telescope Compact Array  is part of the Australia Telescope National Facility, which is funded by the Australian Government for operation as a National Facility managed by CSIRO.
 J.F.G. and L.F.M are
partially supported by MINECO (Spain) grant  AYA2014-57369-C3-3 (co-funded by
FEDER). L.U. acknowledges support from grant PE9-1160 of the Greek General
Secretariat for Research and Technology in the framework of the program
Support of Postdoctoral Researchers.






%



\clearpage

\appendix

\section{Parameters of the individual OH components}

In these tables we list the velocity, position, flux density, and polarisation fraction of the individual spectral features that we could identify.
We only list features higher than 2 sigma over several contiguous channels. Some components are clearly detected in the spectra, but they are too weak to obtain meaningful values of spatial location and polarisation fractions and therefore, these values are left blank in their corresponding table entry.

\begin{table*}
	\centering
	\caption{OH components in NGC 6302 at 1612 MHz.}
	\label{tab:ngc_1612}
	\begin{tabular}{lllllll} %
		\hline
		$V_{\rm LSRK}$ & R.A.(J2000) & Dec(J2000) &  Error & $S_\nu$ & $p_c$ & $p_l$\\
		km s$^{-1}$	  &				& 			& (arcsec)	& (Jy) \\
		\hline
		$-51.4$		  & 17:13:44.59 & $-37$:06:15.8 & (0.3, 0.5) & $0.103\pm 0.024$ & $<0.22$ & $<0.4$\\
		$-41.5$		  & 17:13:44.330 & $-37$:06:10.7 & (0.17, 0.3) & $0.586\pm 0.026$ &$<0.04$ & $<0.07$\\
		\hline
	\end{tabular}
\end{table*}

\begin{table*}
	\centering
	\caption{OH components in JaSt 23.}
	\label{tab:jast_masers}
	\begin{tabular}{lrllllll} %
		\hline
		line & $V_{\rm LSRK}$ & R.A.(J2000) & Dec(J2000) &  Error & $S_\nu$ & $p_c$ & $p_l$\\
		(MHz) & (km s$^{-1}$)	  &				& 			& (arcsec)	& (Jy) \\
		\hline
		1612 & $110.1$ & 17:40:23.077 & $-27$:49:12.0 & (0.10, 0.3) & $0.87\pm 0.03$ & $<0.022$ & $<0.04$\\
		1612 & $112.8$ & 17:40:23.067 & $-27$:49:12.0 & (0.11, 0.3) & $1.19\pm 0.03$ & $<0.016$ & $<0.03$\\
		1612 & $115.7$ & 17:40:23.070 & $-27$:49:12.07 & (0.08, 0.23) & $1.36\pm 0.04$ & $<0.014$ & $<0.022$\\
		1612 & $119.3$ & 17:40:23.075 & $-27$:49:12.2 & (0.13, 0.4) & $1.05\pm 0.03$ & $0.081\pm 0.014$ & $<0.03$\\
		1665 & $115.3$ & 17:40:23.038 & $-27$:49:12.9 & (0.10, 0.3) & $0.350\pm 0.023$ & $<0.05$ & $<0.09$\\
		1665 & $118.2$ & 17:40:23.066 & $-27$:49:12.4 & (0.15, 0.4) & $0.30\pm 0.03$ & $0.13\pm 0.05$ & $<0.11$\\
		\hline
	\end{tabular}
\end{table*}

\begin{table*}
	\centering
	\caption{OH components in IRAS 17375-2759. Corrected by primary beam}
	\label{tab:iras0_masers}
	\begin{tabular}{lrllllll} %
		\hline
		line & $V_{\rm LSRK}$ & R.A.(J2000) & Dec(J2000) &  Error & $S_\nu$ & $p_c$ & $p_l$\\
		(MHz) & (km s$^{-1}$)	  &				& 			& (arcsec)	& (Jy) \\
		\hline
		1612 & $+22.5$ & 17:40:38.575 & -28:01:02.5 &	(0.18, 0.5)	& $0.82\pm 0.04$ & $<0.023$ & $<0.04$\\
		1612 & $+24.1$ & 17:40:38.574 &  -28:01:02.3 & (0.17, 0.5) & $0.87\pm 0.04$ & $<0.022$ & $<0.03$\\
		1612 & $+30.4$ & 17:40:38.52 & -28:01:03 & (0.9, 3) & $0.20\pm 0.04$ & $<0.09$ & $<0.15$\\
		1667 & $+8.6$	&			&			&			& $0.04\pm 0.04$\\
		1667 & $+19.1$ & 17:40:38.59 &	-28:01:02.1	& (0.3, 0.9) & $0.38\pm 0.04$ & $<0.05$ & $<0.11$\\
		\hline
	\end{tabular}
\end{table*}

\begin{table*}
	\centering
	\caption{OH components in IRAS 17393-2727.}
	\label{tab:iras_masers}
	\begin{tabular}{lrlllllll} %
		\hline
		line & $V_{\rm LSRK}$ & R.A.(J2000) & Dec(J2000) &  Error & $S_\nu$ & $p_c$ & $p_l$ &  $\psi$\\
		(MHz) & (km s$^{-1}$)	  &				& 			& (arcsec)	& (Jy) & 	& 		& (deg)\\
		\hline
		1612 & $-123.2$ & 17:42:33.14	& -27:28:25.5	& (0.3, 0.3) & $73.7  \pm 1.2$ & $0.0370\pm 0.0007$ & $0.0411\pm 0.0006$	& $-33.2\pm 0.6$\\ 
		1612 & $-122.4$ & 17:42:33.13  & -27:28:25.4 & (0.4, 0.4) & $52.1\pm 1.2$ & $0.0094\pm 0.0009$ & 	$0.0155\pm 0.0008$ 	& $+37.6\pm 2.2$ \\
		1612 & $-120.7$ & 17:42:33.136 & -27:28:25.4 & (0.23, 0.3) &  $37.9\pm 0.4$ & $0.1039\pm 0.0013$ & $0.0255\pm 0.0011$	& $+10.2\pm 1.9$ \\
		1612 & $-117.7$ & 17:42:33.14 & -27:28:25.6 & (0.3, 0.3) & $5.07\pm 0.10$ & $0.097\pm 0.010$ & $0.130\pm 0.008$ 	& $-12.42\pm 0.05$    \\
		1612 & $-114.8$ & 17:42:33.15 & -27:28:25.8 & (0.4, 0.5) & $3.17\pm 0.08$ & $0.363\pm 0.020$ & $0.062\pm 0.014$	& $+40.58\pm 0.08$\\
		1612 & $-111.5$ & 17:42:33.14 & -27:28:25.5 & (0.8, 1.0) & $0.88\pm 0.04$ & $0.14\pm 0.06$ & $0.08\pm 0.05$	& $-39.8\pm 0.5$\\
		1612 & $-93.4$ & 17:42:33.13 & -27:28:25.6 & (0.7, 0.8) & $1.38\pm 0.04$ & $0.08\pm 0.04$ & $0.05\pm 0.03$ 	& $-1.1\pm 0.5$\\
		1665 & $-124.2$ & 17:42:33.10 & -27:28:25.1 & (2.0, 2.4)& $0.32\pm 0.04$ & $0.25\pm 0.14$ & $<0.3$ \\
		1665 & $-116.4$ & 17:42:33.13 & -27:28:25.6 & (0.3, 0.4) & $0.85\pm 0.05$ & $0.26\pm 0.03$ & $<0.06$\\
		1665 & $-112.7$ & 17:42:33.2 & -27:28:26 & (3, 3) & $0.19\pm 0.05$ & $<0.16$ & $<0.3$\\
		1665 & $-111.0$ & 17:42:33.1 & -27:28:27 & (5, 6) & $0.14\pm 0.05$\\
		1665 & $-103.0$ & 17:42:33.1 & -27:28:24 & (5, 5) & $0.10\pm 0.05$\\
		1667 & $-126.4$ & 17:42:33.16 & -27:28:25.5 & (1.1, 1.3) & $0.39\pm 0.04$ & $0.11\pm 0.05$ & $<0.4$\\
		1667 & $-124.5$ & 17:42:33.16 & -27:28:26.0 & (0.8, 0.9) & $1.23 \pm 0.05$ & $0.13\pm 0.02$ &  $0.18\pm 0.04$ & $+41.24\pm 0.17$\\
		1667 & $-122.3$ & 17:42:33.17 & -27:28:25.7 & (2.0, 2.0) & $0.18\pm 0.04$ & $<0.3$ & $<0.5$\\
		1667 & $-120.0$ & 17:42:33.12 & -27:28:25.5 & (0.8, 1.0) & $0.41\pm 0.04$ & $0.15\pm 0.03$ & $<0.13$\\
		1667 & wide$^a$ & 17:42:33.15 & -27:28:25.5 & (0.5, 0.6) & $\simeq 0.14$ & $<0.23$ & $<0.4$\\
		\hline
	\end{tabular}
	
	$^a$Wide component, from $\simeq -145$ to $-130$ km s$^{-1}$
\end{table*}

\begin{table*}
	\centering
	\caption{OH components in Vy 2-2.}
	\label{tab:vy_masers}
	\begin{tabular}{lrlllllll} %
		\hline
		line & $V_{\rm LSRK}$ & R.A.(J2000) & Dec(J2000) &  Error & $S_\nu$ & $p_c$ & $p_l$ & $\psi$\\\
		(MHz) & (km s$^{-1}$)	  &				& 			& (arcsec)	& (Jy) & & & (deg)\\
		\hline
		1612 & $-61.9$ & 19:24:22.183	& 	+09:53:56.5	& (0.07, 0.5) &$4.33  \pm 0.06$ & $0.013\pm 0.006$ & $0.022\pm 0.013$ & $+18.9\pm 0.4$\\
		1720 & $-54.8$ & 19:24:22.21 &  +09:53:57 & (0.9, 6) & $0.15\pm 0.04$ & $<0.3$ & $<0.5$\\
		\hline
	\end{tabular}	
\end{table*}

\begin{table*}
	\centering
	\caption{OH components in K 3-35.}
	\label{tab:k3_masers}
	\begin{tabular}{lrllllll} %
		\hline
		line & $V_{\rm LSRK}$ & R.A.(J2000) & Dec(J2000) &  Error & $S_\nu$ & $p_c$ & $p_l$\\
		(MHz) & (km s$^{-1}$)	  &				& 			& (arcsec)	& (Jy) \\
		\hline
		1612 & $-4.8$ &				&			&			& $0.06\pm 0.03$\\
		1612 & $-2.7$ & 19:27:44.11 & +21:30:05 & (1.1, 4) & $0.10\pm 0.03$ \\
		1612 & $+9.2$ & 19:27:44.03 & +21:30:04.4 & (0.3, 1.7) & $0.87\pm 0.03$ & $<0.05$ & $<0.08$\\
		1612 & $+18.9$ & 19:27:44.02 & +21:30:03.0 & (0.4, 1.7) & $0.33\pm 0.03$ & $<0.14$ & $<0.22$\\
		1612 & $+21.8$ & 19:27:44.02 & +21:30:02 & (1.4, 6) & $0.20\pm 0.04$ & $0.72\pm 0.25$ & $<0.3$\\
		1665 & $+17.4$ &				&		&			& $0.08\pm 0.03$\\
		1665 & $+22.1$ &				&		&			& $0.07\pm 0.04$\\
		1720 & $+21.2$ & 19:27:44.02 & +21:30:00.9 & (0.4,1.7) & $0.21\pm 0.04$ & $0.50\pm 0.23$ & $<0.4$\\
		\hline
	\end{tabular}
\end{table*}


\bsp	
\label{lastpage}
\end{document}